\documentclass[twocolumn,10pt,journal,a4paper]{IEEEtran}

\usepackage{times,amsmath,epsfig}

\input epsf
\title{\huge  Adaptive Frequency Domain Detectors for SC-FDE in Multiuser DS-UWB Systems with Structured Channel Estimation and Direct Adaptation}

\author{
{Sheng Li ~and~ Rodrigo C. de Lamare}
\thanks{This work is supported by the Department of Electronics, University of York. The authors are with the Communications Research
Group, Department of Electronics, University of York, York, YO10
5DD, UK (e-mail: \{sl546,rcdl500\}@ohm.york.ac.uk).}}

\begin{document}
\maketitle
\begin{abstract}
In this paper, we propose two adaptive detection schemes based on
single-carrier frequency domain equalization (SC-FDE) for multiuser
direct-sequence ultra-wideband (DS-UWB) systems, which are termed
structured channel estimation (SCE) and direct adaptation (DA). Both
schemes use the minimum mean square error (MMSE) linear detection
strategy and employ a cyclic prefix. In the SCE scheme, we perform
the adaptive channel estimation in the frequency domain and
implement the despreading in the time domain after the FDE. In this
scheme, the MMSE detection requires the knowledge of the number of
users and the noise variance. For this purpose, we propose simple
algorithms for estimating these parameters. In the DA scheme, the
interference suppression task is fulfilled with only one adaptive
filter in the frequency domain and a new signal expression is
adopted to simplify the design of such a filter. Least-mean squares
(LMS), recursive least squares (RLS) and conjugate gradient (CG)
adaptive algorithms are then developed for both schemes. A
complexity analysis compares the computational complexity of the
proposed algorithms and schemes, and simulation results for the
downlink illustrate their performance.

\end{abstract}

\section{Introduction}
\IEEEPARstart{U}{ltra}-wideband (UWB) technology
\cite{MZwin1998},\cite{LiuqingYang2004} is a promising next
generation short-range wireless communication technique which has
numerous advantages such as potentially very high data rate and low
operation power. The development of UWB communications for
commercial applications has been boosted with the permit to use a
huge (7.5GHz) unlicensed bandwidth that is released by the Federal
Communications Commission (FCC) in the US in 2002
\cite{LiuqingYang2004}-\cite{oppermann2004}. The direct-sequence
ultra-wideband (DS-UWB) communication, spreads the information
symbols with a pseudo-random (PR) code and enables multiuser
communications \cite{oppermann2004}. DS-UWB systems have been
considered as a potential standard physical layer technology for
wireless personal area networks (WPANs) \cite{RFisher2005}.

In multiuser DS-UWB systems, the receiver is required to effectively suppress
the multiple-access interference (MAI) which is caused by the multiuser
communication and the inter-symbol interference (ISI) which is caused by the
multipath channel. In UWB systems, the huge transmission bandwidths introduce a
high degree of diversity at the receiver due to a large number of resolvable
multipath components (MPCs) \cite{dajana2007}. In order to operate in dense
multipath environments with low complexity, single-carrier frequency domain
equalization (SC-FDE) systems with a cyclic prefix have been recently applied
to DS-UWB communications \cite{pk2008}-\cite{simonemorosi2006}. In
\cite{pk2008}, frequency domain minimum mean-square error (MMSE) turbo
equalization scheme is proposed for single-user DS-UWB systems. For multiuser
communications, the frequency domain detector is obtained by combining the
turbo equalizer with a soft interference canceller. In \cite{YWang2009}, the
performance of the linear MMSE detector in SC-FDE and orthogonal
frequency-division multiplexing (OFDM) systems are compared over UWB channels
and the simulation results show that the SC-FDE system is reasonably robust in
the presents of carrier frequency offset and sampling time offset. In
\cite{YWang2006}, a low-complexity channel estimation algorithm is proposed for
single user communication. A new SC block transmission structure was proposed
in \cite{MingXianchang2006}, where a novel despreading scheme was employed in
the frequency domain before channel estimation and equalization. In
\cite{HSato2006}-\cite{simonemorosi2006}, frequency-domain linear multiuser
detection and channel estimation was performed and a linear MMSE equalization
scheme was described. However, in \cite{pk2008}-\cite{simonemorosi2006},
$prior$ knowledge of the channel and the received signal is required and the
parameter estimation problem was not considered in detail.

Adaptive techniques are effective tools for estimating parameters
and are able to deal with channel variations \cite{haykin}. In the
frequency domain, adaptive algorithms are usually more stable and
converge faster than in the time domain \cite{mmorelli2005}. To the
best of our knowledge, these techniques have not been thoroughly
investigated for UWB communications yet. In this work, adaptive
algorithms based on LMS, RLS and CG techniques are developed for
frequency domain detectors in multiuser DS-UWB communications. The
major advantage of the LMS algorithm is its simplicity and this
feature makes the LMS a standard against other linear adaptive
algorithms \cite{haykin}. The RLS algorithm converges faster than
the LMS algorithm but usually requires much higher computation
complexity. The CG method is the most important conjugate direction
(CD) method that is able to generate the direction vectors simply
and iteratively \cite{dgluenberger}. With faster convergence speed
than stochastic gradient techniques and lower complexity than
recursive least squares (RLS) algorithms, CG methods are known as
powerful tools in computational systems
\cite{dpoleary1980}-\cite{ASLalos2008} and hence, suitable for the
DS-UWB communications.

In this work, we present two adaptive detection schemes in the
frequency-domain and apply them to SC-FDE in multiuser DS-UWB
systems. In the first scheme, a structured channel estimation (SCE)
approach that extends \cite{mmorelli2005} to multiuser UWB systems
is carried out separately in the frequency domain and the estimated
channel impulse response (CIR) is substituted into the expression of
the MMSE detector to suppress the ISI. After the frequency domain
processing, the despreading is performed in the time domain to
eliminate the MAI. The LMS and RLS adaptive algorithms for the SCE
with single user SC systems were proposed in \cite{mmorelli2005} and
in this work we extend them to multiuser scenarios. However, the
SCE-RLS has very high complexity because there is an inversion of
matrix that must be computed directly \cite{mmorelli2005}. This
problem motivates us to develop the SCE-CG algorithm, which will be
shown later, has much lower complexity than the SCE-RLS while
performing better than the SCE-LMS and comparable to the SCE-RLS. In
this scheme, the MMSE detector requires the knowledge of the noise
variance and the number of active users. We estimate the noise
variance via the maximum likelihood (ML) method. With a relationship
between the input signal power and the number of users, we propose a
simple and effective approach to estimating the users number. In the
second scheme, which is termed direct adaptation (DA), only one
filter is implemented in the frequency domain to suppress the
interference. It is important to note that with the traditional
signal expression for the multiuser block transmission systems, the
DA scheme requires a matrix-structured adaptive filter in the
frequency domain which leads to prohibitive complex solutions. In
the literature, the adaptive DA scheme in multiuser UWB systems has
not been investigated in detail. Prior work on adaptive frequency
domain algorithms is limited to single-user systems
\cite{waleed2002} and do not exploit the structure created by
multiuser UWB systems with a cyclic prefix. In order to obtain a
simplified filter design, we adopt the signal expression described
in \cite{MingXianchang2006} and extend it into an adaptive parameter
estimation implementation. After obtaining the matrix form of the
MMSE design of such a filter, we convert it into a vector form and
develop LMS, RLS and CG algorithms in the frequency domain that
enables the linear suppression of ISI and MAI. In our proposed DA
scheme, a low complexity RLS algorithm, termed DA-RLS, is obtained
with the new signal expression. The proposed DA-RLS algorithm is
suitable for multiuser block transmission systems. With faster
convergence rate than the DA-LMS and DA-CG, the complexity of the
DA-RLS in the multiuser cases is comparable to the DA-CG. In the
single user scenario, the complexity of the DA-RLS is reduced to the
level of the DA-LMS.
%

The main contributions of this work are listed below.

\begin{itemize}
\item {Two adaptive detection schemes are developed and compared for SC-FDE in multiuser DS-UWB systems. For both schemes, the LMS, RLS and CG algorithms are developed.}
\end{itemize}

\begin{itemize}
\item {In the first scheme, named SCE, adaptive algorithms are developed for estimating the channel coefficients and algorithms for computing the noise variance and the number of active users are also proposed.}
\end{itemize}

\begin{itemize}
\item {In the second scheme, named DA, a new signal model is adopted to enable simplified adaptive implementation. A low-complexity RLS algorithm is then obtained.}
\end{itemize}

\begin{itemize}
\item {The performance and complexity of LMS, RLS and CG algorithms are compared for both schemes.}
\end{itemize}

The rest of this paper is structured as follows. In section
\ref{sec:systemmodel} the system model is detailed. The detection
schemes for the SC-FDE in DS-UWB system are introduced in section
\ref{sec:detectionschemes}. The proposed adaptive algorithms for SCE
and DA schemes are described in section \ref{sec:sceCG} and section
\ref{sec:DACG}, respectively. The complexity analysis for the
adaptive algorithms and the schemes are presented in section
\ref{sec:complexity}. In section \ref{sec:noiseandK}, the approaches
for estimating the noise variance and the number of active users is
detailed. Simulations results of the proposed schemes are shown in
section \ref{sec:simulation} and section \ref{sec:conclusion} draws
the conclusions.

\section{System Model}
\label{sec:systemmodel}

\begin{figure*}[htb]
\begin{minipage}[b]{1.0\linewidth}
  \centering
  \centerline{\epsfig{figure=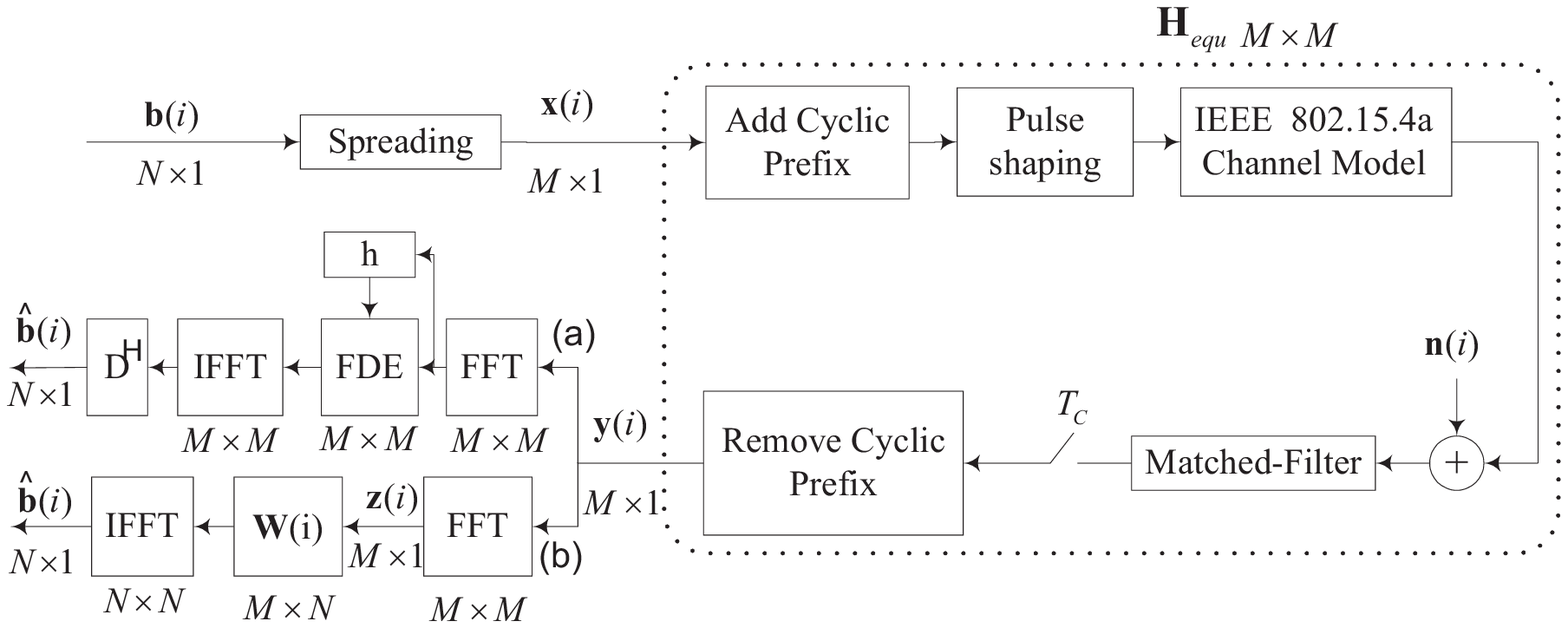,scale=0.8}}
\end{minipage}
\caption{Block diagram of SC-FDE schemes in DS-UWB system, (a) Structured
channel estimation (SCE) and (b) Direct adaptation (DA).} \label{fig:system}
\end{figure*}

In this section, we consider a downlink block-by-block transmission
binary phase-shift keying (BPSK) DS-UWB system with $K$ users. The
block diagram of the system is shown in Fig.\ref{fig:system}. An
$N_{c}$-by-$1$ Walsh spreading code $\mathbf s_{k}$ is assigned to
the $k$-th user. The spreading gain is $N_{c}=T_{s}/T_{c}$, where
$T_{s}$ and $T_{c}$ denote the symbol duration and chip duration,
respectively. At each time instant, an $N$-dimensional data vector
$\mathbf b_{k}(i)$ is transmitted by the $k$-th user, where $N$ is
the block size. We define the signal after spreading as $\mathbf
x_{k}(i)$ and express it in a matrix form as
\begin{equation}
\mathbf x_{k}(i)=\mathbf D_{k}\mathbf b_{k}(i),
\end{equation} where the $M$-by-$N$ ($M=N \times N_{c}$) block diagonal matrix $\mathbf
D_{k}$ is performing the spreading of the data block.

In order to prevent inter block interference (IBI), a
cyclic-prefixed (CP) guard interval is added and the length of the
CP is assumed larger than the CIR. For UWB communications, widely
used pulse shapes include the Gaussian waveforms, Raised-cosine
pulse shaping and Root-Raised Cosine (RRC) pulse shaping
\cite{KazimierzSiwiak2004}. Throughout this paper, the pulse
waveform is modeled as the RRC pulse with a roll-off factor of $0.5$
\cite{RFisher2005}\cite{YWang2006}. With the insertion of the CP at
the transmitter and its removal at the receiver, the Toeplitz
channel matrix could be transformed into an equivalent circulant
channel matrix \cite{YTang2004}. In this work, we adopt the IEEE
802.15.4a standard channel model for the indoor residential non-line
of sight (NLOS) environment \cite{Molisch2005}. This standard
channel model is valid for both low-data-rate and high-data-rate UWB
systems \cite{Molisch2006}. We assume that the timing is perfect and
focus on the channel estimation and interference suppression tasks.
At the receiver, a pulse-matched filter is applied and the received
sequence is then sampled at chip-rate and organized in an
$M$-dimensional vector $\mathbf y(i)$. The equivalent channel is
shown in Fig.\ref{fig:system} and denoted as an $M$-by-$M$ circulant
Toeplitz matrix $\mathbf H_{\rm equ}$, whose first column is
structured with $\mathbf h_{\rm equ}$ zero-padded to length $M$,
where $\mathbf h_{\rm equ}=[h(0), h(1),\dots, h(L-1)]$ is the
equivalent CIR. Hence, the time-domain received signal at the $i$-th
time instant can be expressed as
\begin{equation}
\mathbf y(i)=\sum_{k=1}^{K}\mathbf H_{\rm equ}\mathbf
x_{k}(i)+\mathbf n(i),\label{eq:yi}
\end{equation} where $\mathbf n(i)$ denotes the additive white
Gaussian noise (AWGN).
After the discrete Fourier transform (DFT), the frequency domain
received signal $\mathbf z(i)$ is expressed as
\begin{equation}
\mathbf z(i)=\mathbf F \mathbf y(i),\label{eq:zequalsFy}
\end{equation} where $\mathbf F$ represents the $M$-by-$M$ DFT matrix and
its $(a,b)$-th entry can be expressed as
\begin{equation}
{\rm F}_{a,b}=(1/\sqrt{M}){\rm exp}\{-j(2\pi/M)ab\},
\end{equation} where $a,b$ $\in$ $\{0,M-1\}$.

After the DFT, the frequency domain detectors are implemented to
recover the original signal, as shown in Fig.\ref{fig:system}. We
propose two detection schemes, named SCE and DA, respectively. The
SCE scheme explicitly perform the channel estimation in the
frequency domain, the detection with the estimated channel
coefficients, and finally carry out despreading in the time domain.
The DA scheme implicitly estimates the channel and suppresses the
ISI and MAI together with only one filter and has simpler structure
than the SCE scheme. Without loss of generality, we consider user 1
as the desired user and bypass the subscript of this user for
simplicity.

We define the estimated signal as $\hat {\mathbf b}(i)$ and the
final recovered signal as $\hat {\mathbf b}_{r}(i)$. Hence, for the
SCE scheme, the recovered signal can be expressed as
\begin{equation}
\hat {\mathbf b}_{r}(i)={\rm sign}(\rm {\Re}(\hat {\mathbf
b}(\it{i}))={\rm sign}(\rm {\Re}(\mathbf D^{H} \mathbf F^{H} \mathbf
C^{H} \mathbf {z}(\it{i}))),
\end{equation} where$(\cdot)^H$ denotes the Hermitian transpose,
${\rm sign}(\cdot)$ is the algebraic sign function and $\rm {\Re}
(\cdot)$ represents the real part of a complex number. $\mathbf C$
denotes the frequency domain equalizer. The despreading is denoted
as $\mathbf D^{H}$ which can be considered as the Hermitian
transpose of the spreading matrix. 

For the DA scheme, the final recovered signal can be expressed as
\begin{equation}
\hat {\mathbf b}_{r}(i)={\rm sign}(\rm {\Re}(\hat {\mathbf
b}(\it{i}))={\rm sign}(\rm {\Re}(\mathbf {F}_{\it N}^{H}\mathbf
{W}^{H} \mathbf {z}(\it{i}))),
\end{equation} where $\mathbf {W}$ represents the frequency domain filter that is in an $M$-by-$N$
matrix form. $\mathbf {F}_{\it N}$ is the $N$-by-$N$ DFT matrix. In
this scheme, the channel estimation and the despreading is fulfilled
implicitly together in the filter $\mathbf {W}$.

In the next section, the MMSE designs of the matrix $\mathbf C$ in
the SCE scheme and the $\mathbf {W}$ in the DA scheme will be
detailed.


\section{Proposed Linear MMSE Detection Schemes}
\label{sec:detectionschemes}

In this section, the MMSE design of the proposed schemes will be
detailed. In general, these two detection schemes are based on the
same MMSE problem which aims at minimizing $E[\|\mathbf b(i)-\hat
{\mathbf b}(i)\|^2]$, but they use different approaches to perform
linear detection. For each scheme, some simplifications and
approximations for the later adaptive implementations will also be
presented.

\subsection{Detector with Structured Channel Estimation (SCE)}
The block diagram of the detector with SCE is shown as the branch
(a) in Fig.\ref{fig:system}. Expanding \eqref{eq:zequalsFy}, we have
\begin{equation}
\mathbf z(i)=\mathbf F \mathbf H_{\rm equ}\sum_{k=1}^{K}\mathbf
x_{k}(i)+\mathbf F\mathbf n(i)=\mathbf F\mathbf H_{\rm equ}\mathbf
F^{H}\mathbf F\sum_{k=1}^{K}\mathbf x_{k}(i)+\mathbf F\mathbf n(i),
\label{eq:scezi}
\end{equation}
Bearing in mind the circulant Toeplitz form of the equivalent
channel matrix, we have a diagonal matrix
\begin{equation}
\mathbf{\Lambda}_{\rm H}=\mathbf F\mathbf H_{\rm equ}\mathbf
F^{H},\label{eq:LambdaH}
\end{equation} whose $a$-th diagonal entry can be expressed as
$\tilde{h}_{a}=\sum_{l=0}^{L-1}h_{l}{\rm exp}\{-j(2\pi/M)al\}$. Let
us express it in a more convenient matrix form as
\begin{equation}
\tilde{\mathbf h}=\sqrt{M}\mathbf F_{M,L}\mathbf h_{\rm equ},
\label{eq:tildeh}
\end{equation}where $\tilde{\mathbf h}=[\tilde{h}_{0}, \tilde{h}_{1},\dots,
\tilde{h}_{M-1}]^{T}$ is called frequency domain CIR and $\mathbf
F_{M,L}$ is an $M$-by-$L$ matrix that is structured with the first
$L$ columns of the DFT matrix $\mathbf F$. In order to simplify the
expression of this scheme in later adaptive developments, we include
the constant $\sqrt M$ into the $\mathbf F_{M,L}$, that is
\begin{equation}
\mathbf F_{M,L} \Longleftarrow \sqrt{M}\mathbf F_{M,L}.
\label{eq:fmlfinal}
\end{equation}

The equations \eqref{eq:tildeh} and \eqref{eq:fmlfinal} are
important for the development of the adaptive algorithms in the SCE
scheme which will be detailed later. Here, we develop the MMSE
detector $\mathbf C$ to minimize the following cost function
\begin{equation}
\mathbf {J}_{\rm MSE-SCE}(i)=E[\left\|\mathbf b(i)-\mathbf D^{H}
\mathbf F^{H} \mathbf C^{H} \mathbf {z}(i))\right\|^{2}].
\label{eq:mmsesce}
\end{equation}
Substituting \eqref{eq:scezi} and \eqref{eq:LambdaH} into
\eqref{eq:mmsesce} and assuming that the noise sequence and the
signal sequences are uncorrelated to each other, we can obtain the
expression of the detector as
\begin{equation}
\mathbf C_{\rm{MMSE}}=\left(\mathbf{\Lambda}_{\rm H}\mathbf F
\mathbf D_{\rm all}\mathbf D_{\rm all}^{H}\mathbf
F^{H}\mathbf{\Lambda}_{\rm H}^{H}+\sigma^{2}\mathbf I_{\rm M}
\right)^{-1}\mathbf{\Lambda}_{\rm H}, \label{eq:mmsesce_C}
\end{equation} where the $M$-by-$NK$ block diagonal matrix $\mathbf D_{\rm all}$
contains the information of the spreading codes for all the users
and can be expressed as
\begin{equation}
{\renewcommand{\baselinestretch}{0.3} { \mathbf D_{\rm
all}=\begin{bmatrix}
 \mathbf s_{1}\dots\ \mathbf s_{K}    &                         &               &      \\
                  &   \mathbf s_{1}\dots\ \mathbf s_{K}        &               &       \\
                  &                         &\ddots \\
                  &                         &               &   \mathbf s_{1}\dots\ \mathbf s_{K}    \\
\end{bmatrix}.}}
\label{eq:D_all}
\end{equation}
For the adaptive implementation, the downlink terminal usually dose
not have the information of the spreading codes of other users.
Hence, in this work, we adopted the approximation $\mathbf D_{\rm
all}\mathbf D_{\rm all}^{H}\approx (K/N_{c})\mathbf I_{\rm M}$ for
the development of the adaptive algorithms. This approximation can
also lead to a diagonal MMSE detector that can be considered as a
sub-optimal solution \cite{simonemorosi2006}
\begin{equation}
\hat{\mathbf C}=\left(\frac{K}{N_{c}}\mathbf{\Lambda}_{\rm
H}\mathbf{\Lambda}_{\rm H}^{H}+\sigma_{e}^{2}\mathbf I_{\rm M}
\right)^{-1}\mathbf{\Lambda}_{\rm H}. \label{eq:mmsesce_Capprox}
\end{equation}

From the expression of \eqref{eq:mmsesce_Capprox}, it is clear that
the remaining tasks of the SCE scheme for the adaptive
implementation are to estimate the channel coefficients
$\tilde{\mathbf h}$, the noise variance $\sigma_{e}^2$ and the
number of active users $K$. The proposed algorithms for estimating
these parameters will be presented in later sections.
\subsection{Detector with Direct Adaptation (DA)}
\label{sec:detda}

The block diagram of the DA scheme is shown as the branch (b) in
Fig.\ref{fig:system}. This scheme has much simpler system structure
than the SCE scheme. However, if we go directly with the signal
model used for the SCE scheme, the resulting adaptive filter for DA
schemes will be in an $M$-by-$N$ matrix form which means very high
complexity. Thanks to the new signal model proposed in
\cite{MingXianchang2006}, we can explore the structure of the MMSE
detector in SC-FDE systems more efficiently. In this work, we adopt
this new signal model and extend it to simplify the design of the
adaptive filters. It will be clear soon how the new signal model we
adopted can significantly reduce the complexity of the adaptive
filter implementation in the DA scheme.

Firstly, we can express the transmitted signal from the $k$-th user
as
\begin{equation}
\mathbf x_{k}(i)=\mathbf S_{k}\mathbf b_{k,{\rm e}}(i),
\end{equation} where the $M$-by-$M$ ($M=N \times N_{c}$) spreading matrix $\mathbf S_{k}$
has a circulant Toeplitz form as \cite{MingXianchang2006}
\begin{equation*}
{\renewcommand{\baselinestretch}{0.3} \footnotesize{
 \mathbf
S_{k}=\begin{bmatrix}
 s_{k}(1)     &             &               & s_{k}(2)      \\
 s_{k}(2)     & s_{k}(1)    &               & \vdots        \\
 \vdots       & s_{k}(2)    &               & s_{k}(N_{c})  \\
s_{k}(N_{c})  & \vdots      &\ddots         &              \\
             & s_{k}(N_{c})&\ddots \\
             &             &\ddots \\
              &             &               & s_{k}(1)      \\
\end{bmatrix},}}
\end{equation*}
The equivalent $M$-dimensional expanded data vector is
\begin{equation*}
\mathbf b_{k,{\rm e}}(i)=[b_{k}(1),\mathbf
0_{N_{c}-1},b_{k}(2),\mathbf 0_{N_{c}-1},\cdots,b_{k}(N),\mathbf
0_{N_{c}-1}]^{T},
\end{equation*} where $(\cdot)^T$ is the transpose. Hence, with the
new signal model, the frequency domain received signal becomes
\begin{equation}
\mathbf z(i)=\mathbf F \mathbf y(i)=\sum_{k=1}^{K}\mathbf F \mathbf
H_{\rm equ}\mathbf S_{k}\mathbf b_{k,{\rm e}}(i)+\mathbf F\mathbf
n(i),
\end{equation} Since both $\mathbf H_{\rm equ}$ and $\mathbf S_{k}$ are circulant
Toeplitz matrices, their product also has the circulant Toeplitz
form. This feature makes $\mathbf{\Lambda}_{k}=\mathbf F \mathbf
H_{\rm equ}\mathbf S_{k} \mathbf F^{H}$ a diagonal matrix. Hence, we
have
\begin{equation}
\mathbf z(i)=\sum_{k=1}^{K}\mathbf F \mathbf H_{\rm equ}\mathbf
S_{k} \mathbf F^{H}\mathbf F\mathbf b_{k,{\rm e}}(i)+\mathbf
F\mathbf n(i)=\sum_{k=1}^{K}\mathbf{\Lambda}_{k} \mathbf F\mathbf
b_{k,{\rm e}}(i)+\mathbf F\mathbf n(i).
\end{equation}
We can further expand $\mathbf F\mathbf b_{k,{\rm e}}(i)$ as
\cite{MingXianchang2006}
\begin{equation}
\mathbf F\mathbf b_{k,{\rm e}}(i)=(1/\sqrt {N_{c}})\mathbf I_{\rm
e}\mathbf F_{N}\mathbf b_{k}(i),
\end{equation}
where $\mathbf F_{N}$ denotes the $N$-by-$N$ DFT matrix and the
$M$-by-$N$ matrix $\mathbf I_{\rm e}$ are structured as
\begin{equation}
\mathbf I_{\rm e}=[\underbrace{\mathbf I_{N},\cdots,\mathbf
I_{N}}_{N_{c}}]^{T}. \end{equation} where $\mathbf I_{N}$ denotes
the $N$-by-$N$ identity matrix. Finally, the frequency domain
received signal $\mathbf z(i)$ is expressed as
\begin{equation}
\mathbf z(i)=\sum_{k=1}^{K}(1/\sqrt
{N_{c}})\mathbf{\Lambda}_{k}\mathbf I_{\rm e}\mathbf F_{N}\mathbf
b_{k}(i)+\mathbf F\mathbf n(i). \label{eq:fdinputzi}
\end{equation}
In the DA scheme, an $M$-by-$N$ MMSE filter $\mathbf {W}(i)$ can be
developed via the following cost function:
\begin{equation}
\mathbf {J}_{\rm MSE-DA}(i)=E[\left\|\mathbf b(i)-\mathbf
{F}_{N}^{H}\mathbf W^{H}(i)\mathbf z(i)\right\|^{2}].
\label{eq:mmse}
\end{equation}
The MMSE solution of \eqref{eq:mmse} is
\begin{equation}
\mathbf W_{\rm MMSE}=\mathbf R^{-1}\mathbf P, \label{eq:wmmse}
\end{equation}
where
\begin{equation}
\mathbf R = E[\mathbf z(i)\mathbf
z^{H}(i)]=(1/N_{c})\sum_{k=1}^{K}\mathbf{\Lambda}_{k}\mathbf I_{\rm
e}\mathbf I_{\rm e}^{H}\mathbf{\Lambda}_{k}^{H}+\sigma^{2}\mathbf
I;~~\mathbf P = E[\mathbf z(i)\mathbf b^{H}(i)]=(1/\sqrt
{N_{c}})\mathbf{\Lambda}_{k}\mathbf I_{\rm e}.\label{eq:rpmmse}
\end{equation}
Expanding \eqref{eq:wmmse}, the MMSE solution can be expressed as
\begin{equation}
\begin{split}
\mathbf W_{\rm
MMSE}&=\left(\frac{1}{N_{c}}\sum_{k=1}^{K}\mathbf{\Lambda}_{k}\mathbf
I_{\rm e}\mathbf I_{\rm
e}^{H}\mathbf{\Lambda}_{k}^{H}+\sigma^{2}\mathbf
I\right)^{-1}\frac{\mathbf{\Lambda}_{k}\mathbf I_{\rm e}}{\sqrt
{N_{c}}}=\mathbf V\mathbf I_{\rm e},\\
\end{split}
\end{equation} where the $M$-by-$M$ matrix $\mathbf V$ is
\begin{equation}
\mathbf V=\frac{1}{\sqrt
{N_{c}}}\left(\frac{1}{N_{c}}\sum_{k=1}^{K}\mathbf{\Lambda}_{k}\mathbf
I_{\rm e}\mathbf I_{\rm
e}^{H}\mathbf{\Lambda}_{k}^{H}+\sigma^{2}\mathbf
I\right)^{-1}\mathbf{\Lambda}_{k}.
\end{equation}
Note that the matrix $\mathbf V$ can be expressed as
$N_{c}$-by-$N_{c}$ block matrices $\mathbf v_{ij}$, $i,j$ $\in$
$\{1,N_{c}\}$, each $\mathbf v_{ij}$ is a $N$-by-$N$ diagonal
matrix. Hence, we take a closer look at the product of $\mathbf V$
and $\mathbf I_{\rm e}$:
\begin{equation}
 \mathbf V\mathbf I_{e}=\begin{bmatrix}
 \mathbf v_{1,1}         & \mathbf v_{1,2}        & \dots       &\mathbf v_{1,N_{c}}      \\
 \mathbf v_{2,1}         & \mathbf v_{2,2}        & \dots       &\mathbf v_{2,N_{c}}        \\
 \vdots                  &  \vdots                & \vdots      & \vdots  \\
 \mathbf v_{N_{c},1}     & \mathbf v_{N_{c},2}    & \dots       & \mathbf v_{N_{c},N_{c}}      \\
\end{bmatrix}\begin{bmatrix}
 \mathbf I_{N}  \\
 \mathbf I_{N}  \\
 \vdots         \\
 \mathbf I_{N}  \\
\end{bmatrix}=\begin{bmatrix}
 \sum_{j=1}^{N_{c}}\mathbf v_{1,j}  \\
 \sum_{j=1}^{N_{c}}\mathbf v_{2,j}  \\
 \vdots         \\
 \sum_{j=1}^{N_{c}}\mathbf v_{N_{c},j}  \\
\end{bmatrix}=\begin{bmatrix}
\hat {\mathbf w}_{1}        &              &             &                  \\
                    & \hat {\mathbf w}_{2} &             &                  \\
                    &              &\ddots       &                  \\
                    &              &             & \hat {\mathbf w}_{N_{c}}     \\
\end{bmatrix}\begin{bmatrix}
 \mathbf I_{N}  \\
 \mathbf I_{N}  \\
 \vdots         \\
 \mathbf I_{N}  \\
\end{bmatrix}=\hat {\mathbf W}\mathbf I_{\rm e},
\end{equation} where $\hat {\mathbf w}_{i}=\sum_{j=1}^{N_{c}}\mathbf v_{i,j}$, $i=1,\dots,N_{c}$, are diagonal matrices. Hence, the
product of $\mathbf V$ and $\mathbf I_{\rm e}$ can be converted into
a product of a $M$-by-$M$ ($M=N \times N_{c}$) diagonal matrix $\hat
{\mathbf W}$ and $\mathbf I_{\rm e}$, where the entries of $\hat
{\mathbf W}$ are $\hat {w}_{l}$, $l=1,\dots,M$, equals the sum of
all entries in the $l$-th row of matrix $\mathbf V$. Finally, we
express the MMSE design as
\begin{equation}
\mathbf W_{\rm MMSE}=\hat{\mathbf W}\mathbf I_{\rm e}={\rm
diag}(\hat {\mathbf w}_{\rm e}) \mathbf I_{\rm e},
\label{eq:vectormmse}
\end{equation} where $\hat {\mathbf w}_{\rm e}=(\hat {w}_{1},\hat {w}_{2},\dots,\hat
{w}_{M})$ is an equivalent filter with $M$ taps.

The design of the MMSE filter in DA scheme can be expressed as
either in \eqref{eq:wmmse} or \eqref{eq:vectormmse}. We remark that
the expression shown in \eqref{eq:vectormmse} enable us to design an
$M$-dimensional adaptive filter rather than an $M$-by-$N$ matrix
form adaptive filter to estimate the MMSE solution. This
simplification significantly reduced the complexity of this scheme.

\section{Adaptive Algorithms For SCE}
\label{sec:sceCG}

In this section, we develop the LMS, RLS and CG adaptive algorithms
for the frequency domain channel estimation in multiuser DS-UWB
communications.


\subsection{SCE-LMS}
Substituting \eqref{eq:tildeh} and \eqref{eq:fmlfinal} into
\eqref{eq:scezi} and defining a diagonal matrix $\mathbf X_{\rm
a}(i)=\rm diag[\mathbf F\sum_{k=1}^{K}\mathbf x_{k}(i)]$, the
rearranged frequency domain received signal becomes
\begin{equation}
\mathbf z(i)=\mathbf X_{\rm a}(i)\tilde{\mathbf h}+\mathbf F\mathbf
n(i)=\mathbf X_{\rm a}(i)\mathbf F_{M,L}\mathbf h_{\rm equ}+\mathbf
F\mathbf n(i). \label{eq:scezifinal}
\end{equation}

In the SCE, we take into account the fact that the length of the
equivalent CIR $\mathbf h_{\rm equ}$ is smaller than the received
signal size \cite{mmorelli2005}. For example, we assume that the
DS-UWB channel in the time domain has 100 sample-spaced taps. This
length of the standard channel contains more than 85 percent of the
total energy and can be considered as an upper bound of the channel
length. In the scenario where the received signal has a length of
$M=256$ chips and we assume that each chip was sampled 3 times,
hence the length of the $\mathbf h_{\rm equ}$ is equal to $L=34$
chips that is much smaller than $M$. As shown in \eqref{eq:tildeh},
we can estimate the $L$-dimensional vector $\mathbf h_{\rm equ}$
rather than the $M$-dimensional vector $\tilde {\mathbf h}$. The
SCE-LMS aims at minimizing the MSE cost function
\begin{equation}
\mathbf {J}_{\rm SCE-LMS}(\hat {\mathbf h}_{\rm equ}(i))=E[||\mathbf
z(i)-\mathbf X(i)\mathbf F_{M,L}\hat {\mathbf h}_{\rm
equ}(i)||^{2}], \label{eq:jsceLMS}
\end{equation} where the frequency domain received signal $\mathbf z(i)$ is shown in
\eqref{eq:scezifinal} and $\mathbf X(i)=\rm diag[\mathbf F\mathbf
x(i)]$, $\mathbf x(i)$ is the pilot signal from the desired user.
The gradient of \eqref{eq:jsceLMS} with respect to $\hat {\mathbf
h}_{\rm equ}(i)$ is
\begin{equation}
\mathbf g_{\rm h}(i)=-E[\mathbf F^{H}_{M,L}\mathbf X^{H}(i)\mathbf
z(i)]+E[\mathbf F^{H}_{M,L}\mathbf X^{H}(i)\mathbf X(i)\mathbf
F_{M,L}]\hat {\mathbf h}_{\rm equ}(i),
\end{equation}
This leads to the SCE-LMS algorithm
\begin{equation}
\hat {\mathbf h}_{\rm equ}(i+1)=\hat {\mathbf h}_{\rm
equ}(i)+\mu_{h}\mathbf F^{H}_{M,L}\mathbf X^{H}(i)\mathbf e_{\rm
h}(i),
\end{equation} where $\mathbf e_{\rm h}(i)=\mathbf z(i)-\mathbf X(i)\mathbf F_{M,L}\hat
{\mathbf h}_{\rm equ}(i)$ denotes the $L$-dimensional error vector
and the constant $\mu_{h}$ is the step size of SCE-LMS.

\subsection{SCE-RLS}
The SCE-RLS algorithm is developed to minimize the least squares
(LS) cost function
\begin{equation}
\mathbf {J}_{\rm SCE-RLS}(\hat {\mathbf h}_{\rm
equ}(i))=\sum_{j=1}^{i}\lambda_{h}^{i-j}\left|\left|\mathbf
z(j)-\mathbf X(j)\mathbf F_{M,L}\hat {\mathbf h}_{\rm
equ}(i)\right|\right|^{2}, \label{eq:jsceRLS}
\end{equation} where $\lambda_{h}$ is the forgetting factor.
Computing the gradient of \eqref{eq:jsceRLS} with respect to $\hat
{\mathbf h}_{\rm equ}(i)$ and setting it to zero, the LS solution is
\begin{equation}
{\mathbf h}_{\rm equ,LS}(i)=\mathbf R_{\rm h}^{-1}(i)\mathbf p_{\rm
h}(i)
\end{equation} where $\mathbf R_{\rm h}(i)=
\sum_{j=1}^{i}\mathbf F^{H}_{M,L}\mathbf X^{H}(j)\mathbf X(j)\mathbf
F_{M,L}$ and $\mathbf p_{\rm h}(i)=\sum_{j=1}^{i}\mathbf
F^{H}_{M,L}\mathbf X^{H}(j)\mathbf z(j).$

Note that there is an inversion of an $L$-by-$L$ matrix $\mathbf
R_{\rm h}(i)$ in this solution. The matrix $\mathbf R_{\rm h}(i)$
can be shown in a recursive way as
\begin{equation}
\mathbf R_{\rm h}(i)=\lambda_{h}\mathbf R_{h}(i-1)+ \mathbf
F^{H}_{M,L}\mathbf X^{H}(i)\mathbf X(i)\mathbf F_{M,L}.
\end{equation}
There is no recursive way to simplify the inversion of this matrix
and hence, we apply the adaptation equation shown in
\cite{mmorelli2005}, that is
\begin{equation}
\hat {\mathbf h}_{\rm equ}(i+1)=\hat {\mathbf h}_{\rm equ}(i)+
\mathbf R_{\rm h}^{-1}(i)\mathbf F^{H}_{M,L}\mathbf X^{H}(i)\mathbf
e_{\rm h}(i),
\end{equation} where $\mathbf e_{\rm h}(i)=\mathbf z(i)-\mathbf X(i)\mathbf F_{M,L}\hat
{\mathbf h}_{\rm equ}(i)$ is the $M$-dimensional error vector. For
the $L$-by-$L$ matrix $\mathbf R_{\rm h}(i)$ , computing its inverse
matrix with Gauss-Jordan elimination requires $L^{3}$ of complex
multiplications \cite{matrixalgebra}. This problem makes the SCE-RLS
a high complexity algorithm and for this reason the performance of
RLS algorithm has not been discussed in \cite{mmorelli2005}. For
this paper, our goal is to implement this approach and assess its
performance against the performance of the proposed SCE-CG
algorithm.

\subsection{SCE-CG}
The SCE-CG aims at minimizing the MSE cost function
\begin{equation}
\mathbf {J}_{\rm SCE-CG}(\hat {\mathbf h}_{\rm equ}(i))=E[||\mathbf
z(i)-\mathbf X(i)\mathbf F_{M,L}\hat {\mathbf h}_{\rm
equ}(i)||^{2}], \label{eq:jscecg}
\end{equation} where the frequency domain input signal $\mathbf z(i)$ is shown in
\eqref{eq:scezifinal} and $\mathbf X(i)=\rm diag[\mathbf F\mathbf
x(i)]$, $\mathbf x(i)$ is the pilot signal from the desired user.
The instantaneous estimate of the gradient of \eqref{eq:jscecg} with
respect to $\hat {\mathbf h}_{\rm equ}(i)$ is
\begin{equation}
\hat{\mathbf {g}}_{\rm h}(i)=-\mathbf F^{H}_{M,L}\mathbf
X^{H}(i)\mathbf e_{\rm h}(i),
\end{equation} where $\mathbf e_{\rm h}(i)=\mathbf z(i)-\mathbf X(i)\mathbf F_{M,L}\hat
{\mathbf h}_{\rm equ}(i)$ denotes the error vector. For each input
data vector, a number of iterations is required for the CG method.
Let us denote the iteration index as $c$. For the $(c+1)$-th
iteration, the estimated $\hat {\mathbf h}_{\rm equ}(i)$ is updated
as
\begin{equation}
\hat {\mathbf h}_{\rm equ,c+1}(i)=\hat {\mathbf h}_{\rm
equ,c}(i)+\alpha_{\rm h,c}(i)\mathbf d_{\rm h,c}(i),
\end{equation} where $\alpha_{\rm h,c}(i)$ is the optimum step size and $\mathbf
d_{h,c}(i)$ is the direction vector for the $c$-th iteration. With
the new estimator $\hat {\mathbf h}_{\rm equ,c+1}(i)$, the error
vector is updated as
\begin{equation}
\begin{split}
\mathbf e_{\rm h, c+1}(i)&=\mathbf z(i)-\mathbf X(i)\mathbf
F_{M,L}\hat {\mathbf h}_{\rm equ,c+1}(i)\\
&=\mathbf e_{\rm h, c}(i)-\alpha_{\rm h,c}(i)\mathbf X(i)\mathbf
F_{M,L}\mathbf d_{\rm h,c}(i).\label{eq:ehc1}
\end{split}
\end{equation}
Since the direction vector $\mathbf d_{\rm h,c}(i)$ is orthogonal to
the inverse gradient vector after the $c$-th iteration
\cite{JaeSungLim1996}, we have
\begin{equation}
\mathbf {d}_{\rm h,c}^{H}(i)[-\hat{\mathbf {g}}_{\rm h,c+1}(i)]=0,
\label{eq:dhcghc1}
\end{equation} where $\hat{\mathbf {g}}_{\rm h,c+1}(i)$$=-\mathbf
F^{H}_{M,L}\mathbf X^{H}(i)\mathbf e_{\rm h,c+1}(i)$.

Substituting \eqref{eq:ehc1} into \eqref{eq:dhcghc1}, we obtain the
expression for the optimum step size
\begin{equation}
\alpha_{\rm h,c}(i)=\frac{-\mathbf d_{\rm h,c}^{H}(i)\hat{\mathbf
{g}}_{\rm h,c}(i)}{\mathbf d_{\rm h,c}^{H}\mathbf F^{H}_{M,L}\mathbf
X^{H}(i)\mathbf X(i)\mathbf F_{M,L}\mathbf d_{\rm h,c}(i)}.
\end{equation}
In the CG methods, the direction vector for each iteration can be
obtained by
\begin{equation}
\mathbf d_{\rm h,c+1}(i)= -\hat{\mathbf {g}}_{\rm
h,c+1}(i)+\beta_{\rm h,c}\mathbf d_{\rm h,c}(i),
\label{eq:scecgdirectv1}
\end{equation} where the constant $\beta_{\rm h,c}$ is determined to fulfill the
convergence requirement for the direction vectors that these vectors
are mutually
conjugate\cite{dpoleary1980},\cite{JaeSungLim1996},\cite{ASLalos2008}.
We adopt the expression for $\beta_{\rm h,c}$ as in
\cite{JaeSungLim1996}
\begin{equation}
\beta_{\rm h,c}=\frac{\hat{\mathbf {g}}_{\rm
h,c+1}^{H}(i)\hat{\mathbf {g}}_{\rm h,c+1}(i)}{-\mathbf d_{\rm
h,c}^{H}(i)\hat{\mathbf {g}}_{\rm h,c}(i)}. \label{eq:bcgbetac11}
\end{equation}
Substituting \eqref{eq:scecgdirectv1} into the term $\mathbf d_{\rm
h,c}^{H}(i)\hat{\mathbf {g}}_{\rm h,c}(i)$ in \eqref{eq:bcgbetac11}
and taking the conjugate feature of the direction vectors into
account, that is $\mathbf d_{\rm h,c-1}^{H}(i)\hat{\mathbf {g}}_{\rm
h,c}(i)=0$, we can find that
\begin{equation}
-\mathbf d_{\rm h,c}^{H}(i)\hat{\mathbf {g}}_{\rm
h,c}(i)=\hat{\mathbf {g}}_{\rm h,c}^{H}(i)\hat{\mathbf {g}}_{\rm
h,c}(i).\label{eq:equ37}
\end{equation}
We remark that the relationship obtained in \eqref{eq:equ37} can
reduce the complexity of the SCE-CG algorithm by $\mathcal{O}(\it {
cL})$, where $c$ is the number of iterations and $L$ is the length
of the equivalent CIR. This is because we have to compute the scalar
term $\hat{\mathbf {g}}_{\rm h,c+1}^{H}(i)\hat{\mathbf {g}}_{\rm
h,c+1}(i)$ in \eqref{eq:bcgbetac11} for the $c$-th iteration.
However, with the relationship shown in \eqref{eq:equ37}, this
scalar term can be used directly in the $(c+1)$-th iteration to save
the computation for the scalar term $-\mathbf d_{\rm
h,c+1}^{H}(i)\hat{\mathbf {g}}_{\rm h,c+1}(i)$.

For the SCE scheme, the CG algorithm has lower computational
complexity than the RLS algorithm while performing better than the
LMS algorithm.

The proposed adaptive algorithms for the SCE scheme are summarized
in the first column of Table \ref{tab:algorithms}.
\begin{table*}
\centering
 \caption{\normalsize Adaptive Algorithms For The Proposed Detection Schemes} {
\begin{tabular}{l|l}
\hline \\
\large {SCE-Scheme} & \large {DA-Scheme}\\
\hline \\
\bfseries {1. Initialization:} & \bfseries {1. Initialization:} \rule{0pt}{2.6ex} \\
{$\hat {\mathbf h}_{\rm equ}(1)=$ }$L$-by-$1$ zero-vector  & {$\hat {\mathbf w}(1)=$ }$M$-by-$1$ zero-vector \\
{For \quad $i=1,2,\dots$}& { For \quad $i=1,2,\dots$}\\
\hline\\
\bfseries {2.1 SCE-LMS} & \bfseries {2.1 DA-LMS}\\
~~~~~$\mathbf e_{\rm h}(i)=\mathbf z(i)-\mathbf X(i)\mathbf F_{M,L}\hat {\mathbf h}_{\rm equ}(i)$& ~~~~~$\mathbf e_{\rm w}(i)=\mathbf b(i)-\mathbf Y(i)\hat{\mathbf w}(i)$\\
~~~~~$\hat {\mathbf h}_{\rm equ}(i+1)=\hat {\mathbf h}_{\rm equ}(i)+\mu_{h}\mathbf F^{H}_{M,L}\mathbf X^{H}(i)\mathbf e_{\rm h}(i)$&~~~~~$\hat{\mathbf w}(i+1)=\hat{\mathbf w}(i)+\mu_{\rm w}\mathbf Y^{H}(i) \mathbf e_{\rm w}(i)$\\
\hline\\
\bfseries {2.2 SCE-RLS} & \bfseries {2.2 DA-RLS}\\
~~~~~$\mathbf R_{\rm h}(i)=\lambda_{h}\mathbf R_{\rm h}(i-1)+ \mathbf F^{H}_{M,L}\mathbf X^{H}(i)\mathbf X(i)\mathbf F_{M,L}$&~~~~~$\mathbf R_{\rm w}(i)=\lambda_{w}\mathbf R_{\rm w}(i-1)+\mathbf Y^{H}(i)\mathbf Y(i)$\\
~~~~~$\mathbf e_{\rm h}(i)=\mathbf z(i)-\mathbf X(i)\mathbf F_{M,L}\hat {\mathbf h}_{\rm equ}(i)$&~~~~~$\mathbf e_{\rm aw}(i)=\mathbf b(i)-\mathbf Y(i)\hat{\mathbf w}(i)$\\
~~~~~$\hat {\mathbf h}_{\rm equ}(i+1)=\hat {\mathbf h}_{\rm equ}(i)+\mathbf R_{\rm h}^{-1}(i)\mathbf F^{H}_{M,L}\mathbf X^{H}(i)\mathbf e_{\rm h}(i)$&~~~~~$\hat{\mathbf w}(i+1)=\hat{\mathbf w}(i)+\mathbf R_{\rm w}^{-1}(i)\mathbf Y^{H}(i)\mathbf e_{\rm aw}(i).$\\
\hline\\
\bfseries {2.3 SCE-CG} & \bfseries {2.3 DA-CG}\\
~~~~~\bfseries {STEP 1: Initialization for iterations}&~~~~~\bfseries {STEP 1: Initialization for iterations} \\
~~~~~$\hat {\mathbf h}_{\rm equ,0}(i)=\hat {\mathbf h}_{\rm equ}(i)$, &~~~~~$\hat {\mathbf w}_{0}(i)=\hat {\mathbf w}(i)$, \\
~~~~~$\mathbf e_{\rm h,0}(i)=\mathbf z(i)-\mathbf X(i)\mathbf F_{M,L}\hat{\mathbf h}_{\rm equ,0}(i)$,&~~~~~$\mathbf e_{\rm w,0}(i)=\mathbf b(i)-\mathbf Y(i)\hat{\mathbf w}_{0}(i)$,\\
~~~~~$\mathbf d_{\rm h,0}(i)= -\hat{\mathbf {g}}_{\rm h,0}(i)=\mathbf F^{H}_{M,L}\mathbf X^{H}(i)\mathbf e_{\rm h,0}(i)$.&~~~~~$\mathbf d_{\rm w,0}(i)= -\hat{\mathbf {g}}_{\rm w,0}(i)=\mathbf Y^{H}(i)\mathbf e_{\rm w,0}(i)$.\\
~~~~~For \quad $c=0,1,2,\dots,(c_{\rm max}-1)$&~~~~~For \quad $c=0,1,2,\dots,(c_{max}-1)$\\
~~~~~\bfseries {STEP 2: Update the channel estimation:}&~~~~~\bfseries {STEP 2: Update the filter weights:}\rule{0pt}{2.6ex}\\
~~~~~$\alpha_{\rm h,c}(i)=\frac{\hat{\mathbf {g}}_{\rm h,c}^{H}(i)\hat{\mathbf {g}}_{\rm h,c}(i)}{\mathbf d_{\rm h,c}^{H}\mathbf F^{H}_{M,L}\mathbf X^{H}(i)\mathbf X(i)\mathbf F_{M,L}\mathbf d_{\rm h,c}(i)},$ &~~~~~$\alpha_{\rm w,c}(i)=\frac{\hat{\mathbf g}_{\rm w,c}^{H}(i)\hat{\mathbf {g}}_{\rm w,c}(i)}{\mathbf d_{\rm w,c}^{H}\mathbf Y^{H}(i)\mathbf Y(i)\mathbf d_{\rm w,c}(i)},$ \\
~~~~~$\hat {\mathbf h}_{\rm equ,c+1}(i)=\hat {\mathbf h}_{\rm equ,c}(i)+\alpha_{\rm h,c}(i)\mathbf d_{\rm h,c}(i),$&~~~~~$\hat{\mathbf w}_{c+1}(i)=\hat{\mathbf w}_{c}(i)+\alpha_{\rm w,c}(i)\mathbf d_{\rm w,c}(i),$\\
~~~~~$\mathbf e_{\rm h, c+1}(i)=\mathbf e_{\rm h, c}(i)-\alpha_{\rm h,c}(i)\mathbf X(i)\mathbf F_{M,L}\mathbf d_{\rm h,c}(i)$,&~~~~~$\mathbf e_{\rm w,c+1}(i)=\mathbf e_{\rm w,c}(i)-\alpha_{\rm w,c}(i)\mathbf Y(i)\mathbf d_{\rm w,c}(i)$,\\
~~~~~$\hat{\mathbf {g}}_{\rm h,c+1}(i)=-\mathbf F^{H}_{M,L}\mathbf X^{H}(i)\mathbf e_{\rm h,c+1}(i)$.&~~~~~$\hat{\mathbf {g}}_{\rm w,c+1}(i)=-\mathbf Y^{H}(i)\mathbf e_{\rm w,c+1}(i)$.\\
~~~~~\bfseries {STEP 3: Adapt the direction vector:}&~~~~~\bfseries {STEP 3: Adapt the direction vector:}\rule{0pt}{2.6ex}\\
~~~~~$\beta_{\rm h,c}=\frac{\hat{\mathbf {g}}_{\rm h,c+1}^{H}(i)\hat{\mathbf {g}}_{\rm h,c+1}(i)}{\hat{\mathbf {g}}_{\rm h,c}^{H}(i)\hat{\mathbf {g}}_{\rm h,c}(i)},$ &~~~~~$\beta_{\rm w,c}=\frac{\hat{\mathbf {g}}_{\rm w,c+1}^{H}(i)\hat{\mathbf {g}}_{\rm w,c+1}(i)}{\hat{\mathbf g}_{\rm w,c}^{H}(i)\hat{\mathbf {g}}_{\rm w,c}(i)},$\\
~~~~~$\mathbf d_{\rm h,c+1}(i)= -\hat{\mathbf {g}}_{\rm h,c+1}(i)+\beta_{\rm h,c}\mathbf d_{\rm h,c}(i)$.&~~~~~$\mathbf d_{\rm w,c+1}(i)= -\hat{\mathbf {g}}_{\rm w,c+1}(i)+\beta_{\rm w,c}\mathbf d_{\rm w,c}(i)$.\\
\\
~~~~~$\hat {\mathbf h}_{\rm equ}(i+1)=\hat {\mathbf h}_{\rm equ,c_{max}}(i)$.&~~~~~$\hat{\mathbf w}(i+1)=\hat{\mathbf w}_{\rm c_{max}}(i)$\\
\hline\\
\bfseries {3. Estimate the data vector}&\bfseries {3. Estimate the data vector}\\
$\mathbf{\Lambda}_{\rm H}(i)=\rm{diag}\left(\mathbf
F_{\it{M,L}}\hat{\mathbf h}_{\rm equ}(\it {i})\right)$,\\
$\hat{\mathbf C}(i)=\left(\frac{\hat{K}}{N_{c}}\mathbf{\Lambda}_{\rm H}(i)\mathbf{\Lambda}_{\rm H}^{H}(i)+\hat{\sigma}_{e}^{2}\mathbf I_{\rm M}\right)^{-1}\mathbf{\Lambda}_{\rm H}$&$\hat{\mathbf b}_{r}(i)={\rm sign}(\rm {\Re}(\mathbf Y(\it {i})\hat{\mathbf w}(\it{i})))$.\\
$\hat {\mathbf b}_{r}(i)={\rm sign}(\rm {\Re}(\mathbf D^{H} \mathbf F^{H}\hat{\mathbf C}(\it {i}) \mathbf {z}(\it{i})))$. \\
 \hline
\end{tabular} }
\label{tab:algorithms}
\end{table*}

\section{Adaptive Algorithms For DA}
\label{sec:DACG}

In this section, we develop the LMS, RLS and CG adaptive algorithms
for the DA scheme with the new signal model presented in section
\ref{sec:detda}. For multiuser block transmission systems, these
techniques can be implemented with a simple receiver structure.

\subsection{DA-LMS}
With the expression in \eqref{eq:vectormmse}, we can estimate the
data vector as
\begin{equation}
\hat{\mathbf b}(i)=\mathbf {F}_{N}^{H}\mathbf I_{\rm
e}^{H}\hat{\mathbf W}^{H}(i)\mathbf z(i)=\mathbf {F}_{N}^{H}\mathbf
I_{\rm e}^{H}\hat{\mathbf Z}(i)\hat{\mathbf w}(i),
\end{equation} where $\hat{\mathbf
Z}(i)={\rm diag}(\mathbf z(i))$ and $\hat{\mathbf w}(i)=\hat{\mathbf
w}_{\rm e}^{*}(i)$ is the adaptive filter weight vector. Since
$\mathbf {F}_{N}$ and $\mathbf I_{\rm e}$ are fixed, we consider the
equivalent $N$-by-$M$ received data matrix as
\begin{equation}
\mathbf Y(i)=\mathbf {F}_{N}^{H}\mathbf I_{\rm e}^{H}\hat{\mathbf
Z}(i),\label{eq:nbymreceivedmatrix}
\end{equation}
and express the estimated data vector as
\begin{equation}
\hat{\mathbf b}(i)=\mathbf Y(i)\hat{\mathbf w}(i).
\end{equation}
Hence, the cost function for developing the DA-LMS algorithm can be
expressed as
\begin{equation}
\mathbf {J}_{\rm DA-LMS}(\hat{\mathbf w}(i))=E[||\mathbf
b(i)-\mathbf Y(i)\hat{\mathbf w}(i)||^{2}]. \label{eq:daLMSMSE}
\end{equation}
The gradient of \eqref{eq:daLMSMSE} with respect to $\hat{\mathbf
w}(i)$ is
\begin{equation*}
\mathbf {g}_{\rm w}(i)= -E[\mathbf Y^{H}(i)\mathbf b(i)]+E[\mathbf
Y^{H}(i)\mathbf Y(i)]{\mathbf w}(i).
\end{equation*}
Using the instantaneous estimates of the expected values in the
gradient, we obtain the DA-LMS as
\begin{equation}
\hat{\mathbf w}(i+1)=\hat{\mathbf w}(i)+\mu_{\rm w}\mathbf Y^{H}(i)
\mathbf e_{\rm w}(i), \label{eq:dalmsw}
\end{equation} where $\mathbf e_{\rm w}(i)=\mathbf b(i)-\mathbf Y(i)\hat{\mathbf
w}(i)$ is the $N$-dimensional error vector and $\mu_{\rm w}$ is the
step size for DA-LMS.

\subsection{DA-RLS}

The DA-RLS algorithm is developed to minimize the least square (LS)
cost function
\begin{equation}
\mathbf {J}_{\rm DA-RLS}(\hat {\mathbf
w}(i))=\sum_{j=1}^{i}\lambda_{w}^{i-j}\big|\big|\mathbf b(j)-\mathbf
Y(j)\hat{\mathbf w}(i)\big|\big|^{2}, \label{eq:jdaRLS}
\end{equation} where $\lambda_{w}$ is the forgetting factor.
Computing the gradient of \eqref{eq:jdaRLS} with respect to $\hat
{\mathbf w}(i)$ and setting it to zero, the LS solution is
\begin{equation}
{\mathbf w}_{\rm LS}(i)=\mathbf R_{\rm w}^{-1}(i)\mathbf p_{\rm
w}(i), \label{eq:daLSw}
\end{equation} where $\mathbf R_{\rm w}(i)=
\sum_{j=1}^{i}\mathbf Y^{H}(j)\mathbf Y(j)$ and $\mathbf p_{\rm
w}(i)=\sum_{j=1}^{i}\mathbf Y^{H}(j)\mathbf b(j).$ We can express
the $M$-by-$M$ (where $M=NN_{c}$) matrix $\mathbf R_{\rm w}(i)$ and
the $M$-dimensional vector $\mathbf p_{\rm w}(i)$ recursively as
\begin{equation}
\mathbf R_{\rm w}(i)=\lambda_{w}\mathbf R_{\rm w}(i-1)+\mathbf
Y^{H}(i)\mathbf Y(i),\label{eq:darlsrw}
\end{equation}
\begin{equation}
\mathbf p_{\rm w}(i)=\lambda_{w}\mathbf p_{\rm w}(i-1)+\mathbf
Y^{H}(i)\mathbf b(i).\label{eq:darlspw}
\end{equation}
With the expression of the received data matrix shown in
\eqref{eq:nbymreceivedmatrix}, we can explore the structure of the
matrix $\mathbf R_{\rm w}(i)$, since
\begin{equation}
\mathbf Y^{H}(i)\mathbf Y(i)=\hat{\mathbf Z}^{H}(i)\mathbf I_{\rm
e}\mathbf {F}_{N}\mathbf {F}_{N}^{H}\mathbf I_{\rm
e}^{H}\hat{\mathbf Z}(i)=\hat{\mathbf Z}^{H}(i)\left(\mathbf I_{\rm
e}\mathbf I_{\rm e}^{H}\right)\hat{\mathbf Z}(i),\label{eq:darlsyhy}
\end{equation} the $M$-by-$M$ sparse matrix $\left(\mathbf I_{\rm
e}\mathbf I_{\rm e}^{H}\right)$ is structured with
$N_{c}$-by-$N_{c}$ block matrices and each block matrix is an
$N$-by-$N$ identity matrix. Bearing in mind that the matrix
$\hat{\mathbf Z}(i)$ is a diagonal matrix, we conclude that $\mathbf
R_{\rm w}(i)$ is an $M$-by-$M$ symmetric sparse matrix which
consists of $N_{c}$-by-$N_{c}$ block matrices and each block matrix
is an $N$-by-$N$ diagonal matrix. The number of nonzero elements in
$\mathbf R_{\rm w}(i)$ equals $MN_{c}$. With Gauss-Jordan
elimination \cite{matrixalgebra}, the inversion of each $N$-by-$N$
diagonal matrix has the complexity $\mathcal{O}(N)$ and the
inversion of $N_{c}$-by-$N_{c}$ such block matrices requires the
complexity $\mathcal{O}(NN_{c}^{3})$, which equals
$\mathcal{O}(MN_{c}^{2})$.
Hence, for the single user case, where $N_{c}=1$, the complexity of
computing $\mathbf R_{\rm w}^{-1}(i)$ is only $\mathcal{O}(M)$. In
addition, equation \eqref{eq:darlsyhy} shows that the complexity of
the recursion to obtain $\mathbf R_{\rm w}(i)$ is low. Since the
matrix $\left(\mathbf I_{\rm e}\mathbf I_{\rm e}^{H}\right)$ can be
pre-stored at the receiver, for each time instant, the computation
complexity to obtain $\mathbf R_{\rm w}(i)$ is only
$\mathcal{O}(MN_{c})$. With these properties, we can investigate a
low complexity RLS algorithm to update the filter vector
recursively.
In order to obtain such a recursion, we apply the method that is
proposed in the appendix B in \cite{mmorelli2005}. We have the
relationship
\begin{equation}
\mathbf R_{\rm w}(i)\hat{\mathbf w}(i+1)=\mathbf p_{\rm
w}(i).\label{eq:fordarlsrwp}
\end{equation} Replacing $\hat{\mathbf w}(i+1)$ with $[\hat{\mathbf w}(i+1)-\hat{\mathbf w}(i)+\hat{\mathbf w}(i)]$ in
\eqref{eq:fordarlsrwp} and using \eqref{eq:darlsrw} and
\eqref{eq:darlspw} obtains
\begin{equation}
\mathbf R_{\rm w}(i)[\hat{\mathbf w}(i+1)-\hat{\mathbf
w}(i)]+[\lambda_{w}\mathbf R_{\rm w}(i-1)+\mathbf Y^{H}(i)\mathbf
Y(i)]\hat{\mathbf w}(i)=\lambda_{w}\mathbf p_{\rm w}(i-1)+\mathbf
Y^{H}(i)\mathbf b(i).\label{eq:darlsderivation1}
\end{equation} Since $\mathbf R_{\rm w}(i-1)\hat{\mathbf w}(i)=\mathbf p_{\rm
w}(i-1)$, \eqref{eq:darlsderivation1} becomes
\begin{equation}
\mathbf R_{\rm w}(i)[\hat{\mathbf w}(i+1)-\hat{\mathbf
w}(i)]=\mathbf Y^{H}(i)\mathbf e_{\rm aw}(i),
\end{equation}where $\mathbf e_{\rm aw}(i)=\mathbf b(i)-\mathbf Y(i)\hat{\mathbf
w}(i)$ is the $N$-dimensional error vector.
Finally, the recursion for updating the filter vector is
\begin{equation}
\hat{\mathbf w}(i+1)=\hat{\mathbf w}(i)+\mathbf R_{\rm
w}^{-1}(i)\mathbf Y^{H}(i)\mathbf e_{\rm aw}(i).\label{eq:darlsw}
\end{equation} We remark that the DA-RLS only consists of \eqref{eq:darlsrw} and
\eqref{eq:darlsw}. The complexity of this algorithm is only
$\mathcal{O}(MN_{c}^{2})$, which is comparable to the DA-CG in
multiuser scenarios and for the single user scenario where
$N_{c}=1$, it reduces to the level of the DA-LMS.

\subsection{DA-CG}

The cost function for developing a CG algorithm in DA scheme can be
expressed as
\begin{equation}
\mathbf {J}_{\rm DA-CG}(\hat{\mathbf w}(i))=E[||\mathbf b(i)-\mathbf
Y(i)\hat{\mathbf w}(i)||^{2}]. \label{eq:dacgMSE}
\end{equation}
The gradient of \eqref{eq:dacgMSE} with respect to $\hat{\mathbf
w}(i)$ is
\begin{equation*}
\mathbf {g}_{\rm w}(i)= -E[\mathbf Y^{H}(i)\mathbf b(i)]+E[\mathbf
Y^{H}(i)\mathbf Y(i)]\hat{\mathbf w}(i).
\end{equation*}
We can use the instantaneous estimates of the expected values and
obtain an estimate of the gradient vector as
\begin{equation}
\hat{\mathbf {g}}_{\rm w}(i)=-\mathbf Y^{H}(i)\mathbf e_{\rm w}(i),
\end{equation} where $\mathbf e_{\rm w}(i)=\mathbf
b(i)-\mathbf Y(i)\hat{\mathbf w}(i)$ is the error vector. Here, we
also define the iteration index as $c$. For the $(c+1)$-th
iteration, the error vector is
\begin{equation}
\mathbf e_{\rm w,c+1}(i)=\mathbf b(i)-\mathbf Y(i)\hat{\mathbf
w}_{\rm c+1}(i),\label{eq:bcge}
\end{equation} where the filter weight vector is updated as
\begin{equation}
\hat{\mathbf w}_{\rm c+1}(i)=\hat{\mathbf w}_{\rm c}(i)+\alpha_{\rm
w,c}(i)\mathbf d_{\rm w,c}(i), \label{eq:bcgw}
\end{equation} where $\mathbf d_{\rm w,c}(i)$ is the direction vector at the
$c$-th iteration. The step size $\alpha_{\rm w,c}(i)$ is determined
to minimize the cost function \eqref{eq:dacgMSE}
\cite{JaeSungLim1996},\cite{ASLalos2008}. Substituting
\eqref{eq:bcgw} in \eqref{eq:bcge}, the error vector can be
expressed as
\begin{equation}
\mathbf e_{\rm w,c+1}(i)=\mathbf e_{\rm w,c}(i)-\alpha_{\rm
w,c}(i)\mathbf Y(i)\mathbf d_{\rm w,c}(i). \label{eq:bcge1}
\end{equation}
Since the direction vector $\mathbf d_{\rm w,c}(i)$ is orthogonal to
the inverse gradient vector after the $c$-th iteration
\cite{JaeSungLim1996}, we have $\mathbf {d}_{\rm
w,c}^{H}(i)[-\hat{\mathbf {g}}_{\rm w,c+1}(i)]=0$, where
$\hat{\mathbf {g}}_{\rm w,c+1}(i)=-\mathbf Y^{H}(i)\mathbf e_{\rm
w,c+1}(i)$. Hence, from \eqref{eq:bcge1}, the optimum step size is
\begin{equation}
\alpha_{\rm w,c}(i)=\frac{-\mathbf d_{\rm w,c}^{H}(i)\hat{\mathbf
{g}}_{\rm w,c}(i)}{\mathbf d_{\rm w,c}^{H}\mathbf Y^{H}(i)\mathbf
Y(i)\mathbf d_{\rm w,c}(i)}.
\end{equation}
The adaptation equation for the direction vector can be expressed as
\begin{equation}
\mathbf d_{\rm w,c+1}(i)= -\hat{\mathbf {g}}_{\rm
w,c+1}(i)+\beta_{\rm w,c}\mathbf d_{\rm w,c}(i),
\label{eq:bcgdirectv1}
\end{equation} where the constant $\beta_{\rm w,c}$ is determined to fulfill the
convergence requirement for the direction vectors that these vectors
are mutually conjugate\cite{dpoleary1980},
\cite{JaeSungLim1996},\cite{ASLalos2008}. We adopt the expression
for $\beta_{\rm w,c}$ as in \cite{JaeSungLim1996}
\begin{equation}
\beta_{\rm w,c}=\frac{\hat{\mathbf {g}}_{\rm
w,c+1}^{H}(i)\hat{\mathbf {g}}_{\rm w,c+1}(i)}{-\mathbf d_{\rm
w,c}^{H}(i)\hat{\mathbf {g}}_{\rm w,c}(i)}, \label{eq:bcgbetac1}
\end{equation}
If we substitute \eqref{eq:bcgdirectv1} into the term $\mathbf
d_{\rm w,c}^{H}(i)\hat{\mathbf {g}}_{\rm w,c}(i)$ in
\eqref{eq:bcgbetac1} and take the conjugate feature of the direction
vectors into account, we can find that
\begin{equation}
-\mathbf d_{\rm w,c}^{H}(i)\hat{\mathbf {g}}_{\rm
w,c}(i)=\hat{\mathbf {g}}_{\rm w,c}^{H}(i)\hat{\mathbf {g}}_{\rm
w,c}(i).\label{eq:equ48}
\end{equation} As explained for \eqref{eq:equ37}, the relationship
obtained in  \eqref{eq:equ48} can save the computational complexity
by $\mathcal{O}(cM)$ for the DA-CG algorithm, where $c$ is the
number of iterations and $M$ is the length of the received signal.

The proposed adaptive algorithms for the DA scheme are summarized in
the second column of Table \ref{tab:algorithms}.
\section{Complexity Analysis}
\label{sec:complexity}
\begin{table}
\centering \caption{\normalsize Complexity analysis}
\begin{tabular}{l l l}
\hline\hline { Algorithm} & Complex Additions & Complex Multiplications\rule{0pt}{2.6ex} \rule[-1.2ex]{0pt}{0pt}\\
\hline
SCE-LMS                & $2ML$                        &$2ML+2M+L$ \rule{0pt}{2.6ex} \\
SCE-RLS                & $2L^{3}+2ML-2L^{2}$          &$2L^{3}+3ML+(2+M)L^{2}$\\
SCE-CG                 & $(2ML+M+3L-3)c$              &$(2ML+4M+4L+1)c$\\
DA-LMS                 & $2MN$                        &$2MN+N$\rule{0pt}{2.6ex}\\
DA-RLS                 & $M(N_{c}^{2}+2N_{c}+2N-2)$   &$M(N_{c}^{2}+6N_{c}+2N-1)$\\
DA-CG                  & $(2MN+2M-2)c$                &$(2MN+2M+N+2)c$\\
\hline
\end{tabular}
\label{tab:Complexity analysis}
\end{table}
In this section, we discuss the complexity of the proposed adaptive
algorithms and the detection schemes.

Table \ref{tab:Complexity analysis} shows the complexity for the
proposed algorithms with respect to the number of complex additions
and complex multiplications for each time instant, where $M$ is the
length of the received signal, $N$ is the length of the data block
and $L$ is the length of the equivalent CIR. For the CG algorithms,
the iteration number is denoted as $c$, which is much smaller than
$M$, say $M=256$, $c=8$. In this work, the complexity of the FFT and
IFFT, which is $\mathcal{O}(M\rm {log_{2}}\it {M})$, is common to
all the techniques and is not shown in this table.

\begin{figure}[htb]
\begin{minipage}[b]{1.0\linewidth}
  \centering
  \centerline{\epsfig{figure=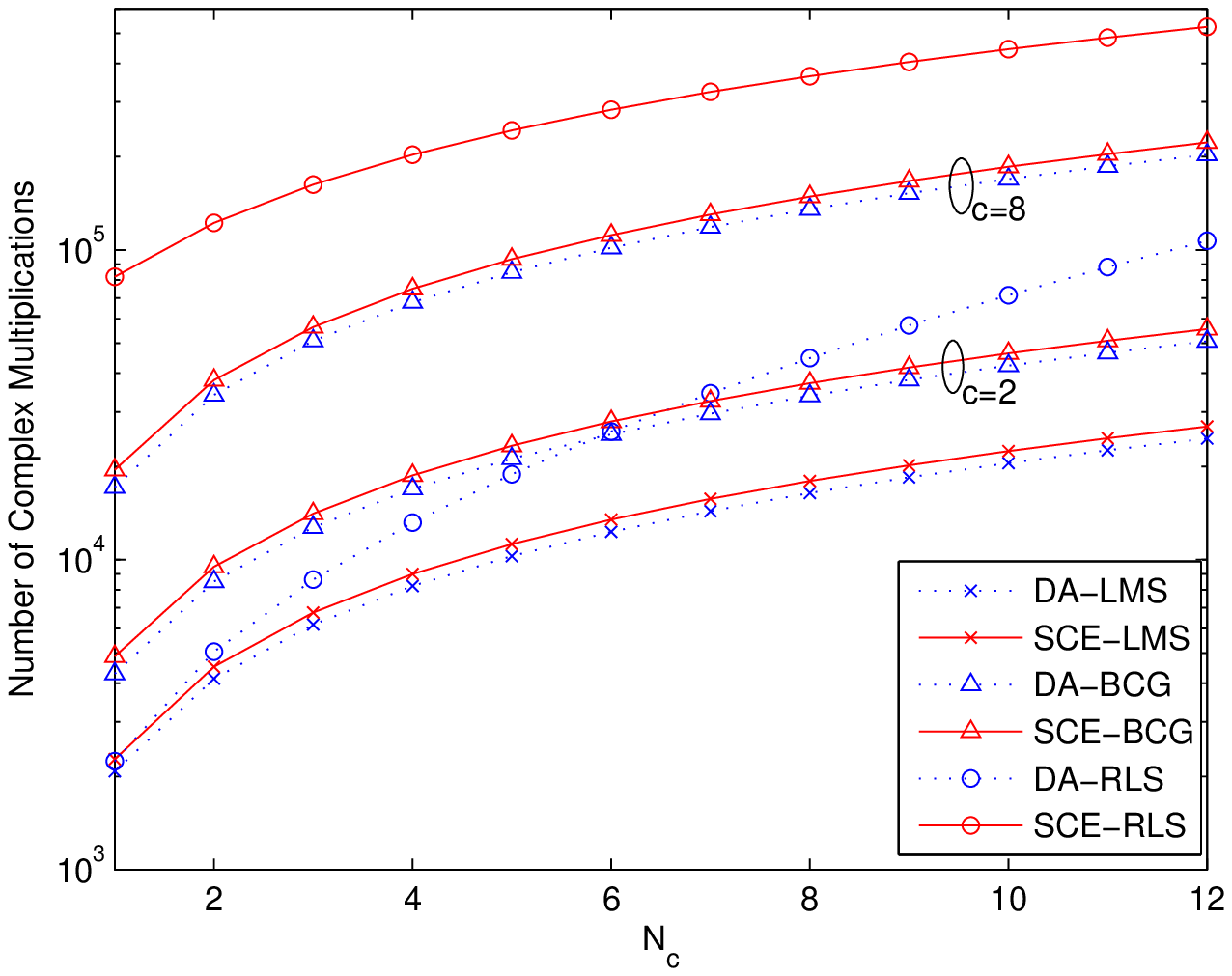,scale=0.8}}
\end{minipage}
\caption{Complexity comparison of the proposed schemes.} \label{fig:complexity}
\end{figure}

It is important to note that for the adaptive algorithms in the SCE
scheme, the complexity is determined by $M$ and $L$, while in the DA
scheme it is determined by $M$ and $N$, bearing in mind that the
spreading gain $N_{c}$ equals $M/N$. Hence, we compare the
complexity of the algorithms with the system parameters that will be
used in the simulation section, say $L=34$ and $N=32$, with
different spreading gain $N_{c}$ (which leads to different received
signal length $M$, since $M=N_{c}N$). Fig. \ref{fig:complexity}
shows the number of complex multiplications for adaptive algorithms
versus different spreading gains. The complexity of the CG
algorithms with iteration number of 2 and 8 are shown in this figure
for comparison. With these system parameters, the SCE-LMS has
similar complexity as DA-LMS, and SCE-CG has similar complexity as
DA-CG. However, the SCE-RLS is the most complex adaptive algorithm
while the DA-RLS has much lower complexity. For the SCE scheme, the
SCE-CG algorithm is significantly simpler than the SCE-RLS. It will
be illustrated by the simulation results that with 8 iterations, the
performance of the SCE-CG algorithm is very close to the SCE-RLS.
For the DA scheme, in the single user scenario where $N_{c}=1$, the
DA-RLS has the same complexity level as the DA-LMS. In the multiuser
case, the complexity of the DA-RLS is comparable to the DA-CG. With
small spreading gains, the DA-RLS has lower complexity than the
DA-CG with only 2 iterations. However, the complexity of the DA-RLS
will be boosted when the spreading gain increases. It will be
illustrated by simulations that the performance of DA-CG is
comparable to the DA-RLS, hence, for multiuser scenarios with
different values of $N_{c}$, the designer can choose either RLS or
CG for the DA scheme.

After discussing the complexity of the adaptive algorithms, let us
consider the complexity of the detection schemes. For the DA scheme,
where only one adaptive filter is implemented and the complexity
shown in Table \ref{tab:Complexity analysis} can be considered as
the whole scheme's complexity. However, for the SCE scheme, the
complexity shown in the table is only for the adaptive channel
estimation. The complexity of estimating the noise variance
$\mathcal{O}(ML^2)$, the number of active users $\mathcal{O}(M)$,
performing the MMSE detection $\mathcal{O}(M)$ and the time domain
despreading $\mathcal{O}(N^2)$ should also be included. Hence, we
conclude that the DA scheme is simpler than the SCE scheme in both
structure and the computational complexity. However, the SCE scheme,
which will be shown later, has better performance than the DA
scheme.

\section{Noise Variance and Number of Active Users Estimation}
\label{sec:noiseandK}

For the SCE scheme, the MMSE detector is generated as
\eqref{eq:mmsesce_Capprox}, which requires the knowledge of the
noise variance $\sigma^{2}_{e}$ and the number of active users $K$.
In this section, we propose an ML estimation algorithm that extends
\cite{mmorelli2005} for estimating $\sigma^{2}_{e}$ in the DS-UWB
system. We consider multiuser communication and the pilot sequence
is generated randomly for each time instant.

The most popular active users number detection schemes for symbol by
symbol transmission systems that are based on the eigenvalue
decomposition have been proposed in
\cite{HSari1995}-\cite{roschmidt1986}. These schemes have very high
complexity and require high SNR to work in our block transmission
system. In this work, we propose a simple approach to estimate the
number of users based on the idea that the power of the received
signal will reflect the number of active users. So firstly, we
develop the relationship between the received signal power, the
noise variance, the estimated channel coefficients and the number of
active users. Then we obtain a simple algorithm to estimate $K$ with
these relationships.
\subsection{Noise Variance Estimation}
Revisiting \eqref{eq:scezifinal}, we have the frequency domain
received signal as
\begin{equation}
\mathbf z(i)=\mathbf X_{\rm a}(i)\mathbf F_{M,L}\mathbf h_{\rm
equ}+\mathbf F\mathbf n(i),
\end{equation} where the diagonal matrix $\mathbf X_{\rm a}(i)=\rm
diag[\sum_{k=1}^{K}\mathbf F\mathbf x_{k}(i)]$. We assume that the
first user is the desired user and define $\mathbf X(i)=\rm
diag[\mathbf F\mathbf x(i)]$. The uncorrelated additive noise is
assumed to be Gaussian distributed with zero mean and variance of
$\sigma^{2}_{e}$. The ML estimator aims at estimating
${\sigma}^{2}_{e}(i)$ and ${\mathbf h}_{\rm equ}(i)$ by maximizing
the log-likelihood function, that is
\begin{equation}
\left[\hat{\sigma}^{2}_{e}(i),\hat{\mathbf h}_{\rm
equ}(i)\right]=\arg \max_{{\sigma}^{2}_{e}(i),{\mathbf
h}_{\rm equ}(i)}\mathbf{\Lambda}({\sigma}^{2}_{e}(i),{\mathbf
h}_{\rm equ}(i)),
\label{eq:loglikelihoodnoiseo1}
\end{equation} where
\begin{equation}
\mathbf{\Lambda}({\sigma}^{2}_{e}(i),{\mathbf
h}_{\rm equ}(i))=-M{\rm ln}({\sigma}^{2}_{e}(i))-\frac{\left\|\mathbf z(i)-\mathbf
B(i){\mathbf h}_{\rm
equ}(i)\right\|^{2}}{{\sigma}^{2}_{e}(i)},
\label{eq:loglikelihoodnoise}
\end{equation} where $\mathbf B(i)=\mathbf X(i)\mathbf F_{M,L}$.
To solve this joint optimization problem, we firstly fix
${\sigma}^{2}_{e}(i)$ and find the optimum $\hat{\mathbf h}_{\rm
equ}(i)$. By calculating the gradient of
\eqref{eq:loglikelihoodnoise} with respect to ${\mathbf h}_{\rm
equ}(i)$ and setting it to zero, we obtain
\begin{equation}
\hat{\mathbf h}_{\rm equ, ML}(i)=\left(\mathbf B^{H}(i)\mathbf
B(i)\right)^{-1}\mathbf B^{H}(i)\mathbf z(i). \label{eq:noiseeh}
\end{equation}
Substituting \eqref{eq:noiseeh} into \eqref{eq:loglikelihoodnoise},
and calculating the gradient of \eqref{eq:loglikelihoodnoise} with
respect to ${\sigma}^{2}_{e}(i)$ and setting it to zero, we obtain
the ML estimate of ${\sigma}^{2}_{e}(i)$
\begin{equation}
\hat{\sigma}^{2}_{e,\rm ML}(i)=\frac{1}{M}\left\|\mathbf
z(i)-\mathbf B(i)\hat{\mathbf h}_{\rm equ,
ML}(i)\right\|^{2}.\label{eq:hatsigma}
\end{equation}
In the training stage of the SCE scheme, we estimate the noise
variance via \eqref{eq:noiseeh} and \eqref{eq:hatsigma}, where the
number of complex multiplications required is
$ML^{2}+L^3+2ML+L^2+M+1$. The cost of this estimator is high and it
is possible to obtain a simplified estimate with the complexity of
$\mathcal{O}(ML)$ by replacing the ML estimate $\hat{\mathbf h}_{\rm
equ, ML}(i)$ with the estimated channel $\hat{\mathbf h}_{\rm
equ}(i)$ that is obtained in section \ref{sec:sceCG}. However, this
will introduce noticeable degradation of the estimation performance
in multiuser scenarios. Since in our SCE scheme, the noise variance
estimator is used for both users number estimation and the MMSE
detection, the degradation of the $\hat{\sigma}^{2}$ will affect the
final performance.

\subsection{Number of Active Users Estimation}

In order to obtain the relationship of the active users number and
the received signal power, let us consider the expected value of the
frequency domain received signal power
\begin{equation}
E[\mathbf z^{H}(i)\mathbf z(i)]=E\left[\left(\mathbf X_{\rm
a}(i)\tilde{\mathbf h}+\mathbf F\mathbf n(i)\right)^{H}\left(\mathbf
X_{\rm a}(i)\tilde{\mathbf h}+\mathbf F\mathbf
n(i)\right)\right]=\tilde{\mathbf h}^{H}E[\mathbf X_{\rm
a}^{H}(i)\mathbf X_{\rm a}(i)]\tilde{\mathbf h}+\sigma_{e}^2 M,
\label{eq:powerofzz}
\end{equation} where $\mathbf z(i)$ is shown in
\eqref{eq:scezifinal}.
Since the $M$-by-$M$ diagonal matrix $\mathbf X_{\rm a}(i)=\rm
diag[\mathbf F\sum_{k=1}^{K}\mathbf x_{k}(\it{i})]$, the $l$-th
entry of the main diagonal can be expressed as
\begin{equation}
\rm X_{\rm a,\it l}(\it{i})=\mathbf F_{\it l}\sum_{k=1}^{K}\mathbf
x_{k}(\it{i}),
\end{equation} where $l=1,2,\dots,M$ and $\mathbf F_{l}$ is the $l$-th row of the DFT
matrix $\mathbf F$. Bearing in mind the fact that $\mathbf F_{l}\mathbf
F^{H}_{l}=1$. Hence, the expected entry in \eqref{eq:powerofzz} can
be expressed as
\begin{equation}
E[\mathbf X_{\rm a}^{H}(i)\mathbf X_{\rm a}(i)]=\rm
diag\left(E[X_{\rm a,\it 1}^2,X_{\rm a,\it 2}^2,\dots,X_{\rm a,\it
M}^2]\right),\label{eq:forKXa}
\end{equation} where
\begin{equation}
E\left[\rm X_{\rm a,\it l}^2\right]=E\left[\mathbf
F_{l}\left(\sum_{k=1}^{K}\mathbf
x_{k}(i)\right)\left(\sum_{k=1}^{K}\mathbf
x_{k}(i)\right)^{H}\mathbf F_{l}^{H}\right]=E\left[\mathbf
F_{l}\mathbf D_{\rm all}\mathbf D_{\rm all}^{H}\mathbf
F_{l}^{H}\right]\approx \frac {K}{N_{c}}, \label{eq:forKXaapprox}
\end{equation} where $\mathbf D_{\rm all}$ is shown in \eqref{eq:D_all} and the approximation used here is the same as in
\eqref{eq:mmsesce_Capprox}, that is $\mathbf D_{\rm all}\mathbf
D_{\rm all}^{H}\approx (K/N_{c})\mathbf I_{\rm M}$.

Finally, substituting \eqref{eq:forKXa} into \eqref{eq:powerofzz}
with the approximation shown in \eqref{eq:forKXaapprox}, we obtain
the relationship which can be expressed as
\begin{equation}
E[\mathbf z^{H}(i)\mathbf z(i)]\approx \frac{K}{N_{c}}\tilde{\mathbf
h}^{H}\tilde{\mathbf h}+\sigma_{e}^2 M,\label{eq:forKfinalideal}
\end{equation} where $K$ is the number of active users,
$\sigma_{e}^2$ is the noise variance and $\tilde{\mathbf h}$ is the
frequency domain channel coefficients. In this work, we have already
obtained the estimator for $\sigma_{e}^2$ and ${\mathbf h}_{\rm
equ}$. The expected received signal power can be estimated via
time-averaging, that is
\begin{equation}
\hat {\rm P}_{\rm r}(i)=\frac {1}{T}\sum_{i=1}^{T}\mathbf
z^{H}(i)\mathbf z(i).
\end{equation}
Hence, the algorithm for estimating $K$ can be expressed as
\begin{equation}
\hat {K}(i)= \left(\hat {\rm P}_{\rm
r}(i)-\hat{\sigma}^{2}_{e}(i)M\right)\frac{N_{c}}{\hat {\rm P}_{\rm
h}(i)},\label{eq:forKfinalused}
\end{equation} where
\begin{equation}
\hat {\rm P}_{\rm h}(i)=(\mathbf F_{M,L}\hat{\mathbf h}_{\rm
equ}(i))^{H}(\mathbf F_{M,L}\hat{\mathbf h}_{\rm equ}(i)).
\end{equation} In order to obtain the integer estimated values, we can set
$\hat {K}(i)$ to the nearest integer towards zero.

We remark that this proposed algorithm is efficient to estimate the
number of active users in the downlink of our block data
transmission system with very low complexity. The only parameter
that is required to compute for this algorithm is the average
received signal power.

\section{Simulation Results}
\label{sec:simulation}

In this section, we apply the proposed SC-FDE schemes and algorithms
to the downlink of a multiuser BPSK DS-UWB system. The pulse shape
adopted in this work is the Root-Raised Cosine (RRC) pulse with the
pulse-width $T_{c}=0.375ns$. The length of the data block is set to
$N=32$ symbols. The Walsh spreading code with a spreading gain
$N_{c}=8$ is generated for the simulations and we assume that the
maximum number of active users is $7$. The channel has been
simulated according to the standard IEEE 802.15.4a channel model for
the NLOS indoor environment as shown in \cite{Molisch2005}. We
assume that the channel is constant during the whole transmission
and the time domain CIR has the length of $100$. This length of the
channel gathered more than 85 percent of the total channel energy of
the standard channel model. The sampling rate of the standard
channel model is $8GHz$. The CP guard interval has the length of
$35$ chips, which has the equivalent length of $105$ samples and it
is enough to eliminate the IBI. The uncoded data rate of the
communication is approximately $293\rm Mbps$. For all the
simulations, the adaptive filters are initialized as null vectors.
This allows a fair comparison between the analyzed techniques of
their convergence performance. In practice, the filters can be
initialized with \textit{prior} knowledge about the spreading code
or the channel to accelerate the convergence. All the curves are
obtained by averaging $100$ independent simulations.

\begin{figure}[htb]
\begin{minipage}[b]{1.0\linewidth}
  \centering
  \centerline{\epsfig{figure=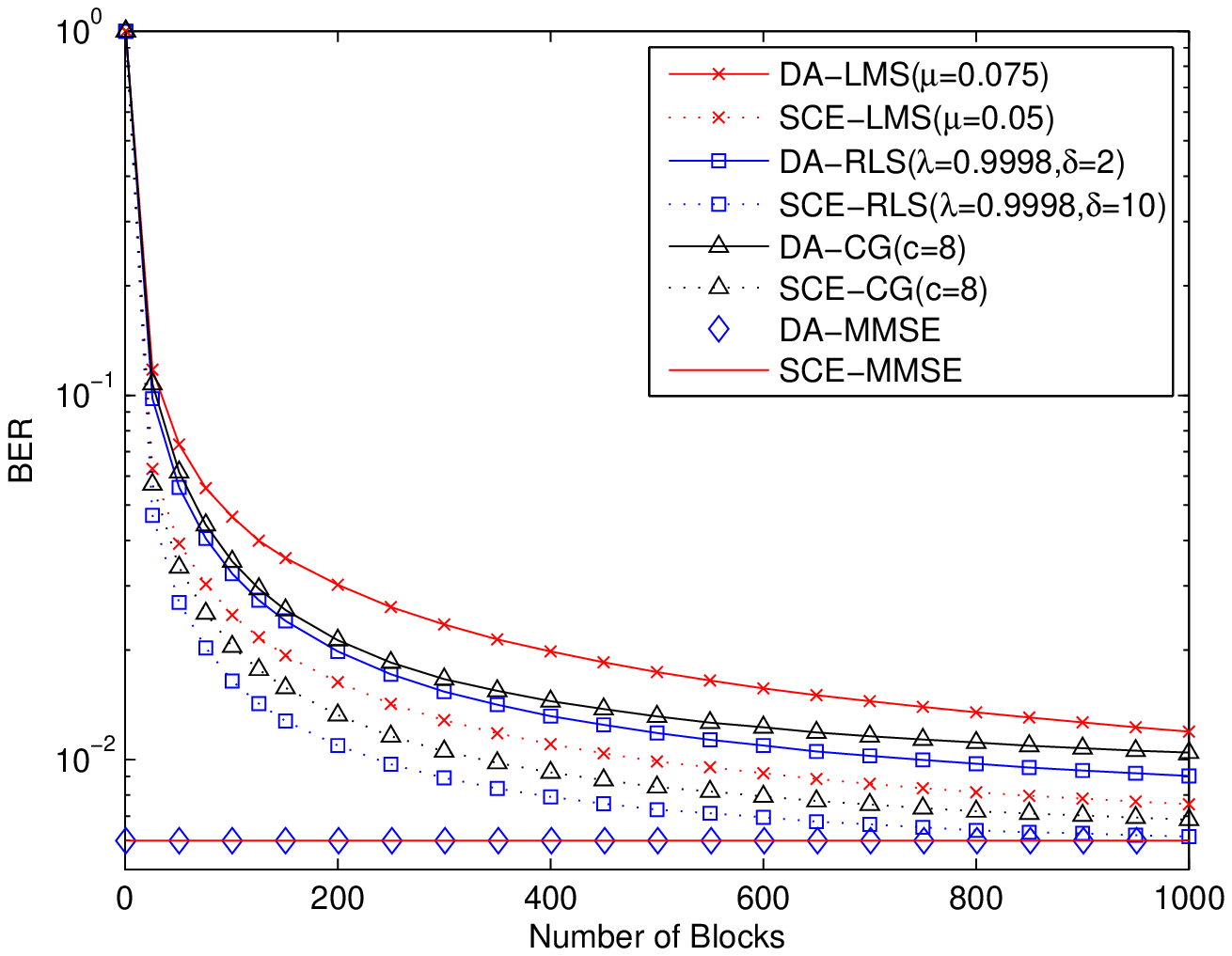,scale=0.8}}
\end{minipage}
\caption{BER performance of the proposed SC-FDE detection schemes versus the
number of training blocks for a SNR=16dB. The number of users is 3.}
\label{fig:16dbblockberperfect}
\end{figure}

The first experiment we perform is to compare the uncoded bit error
rate (BER) performance of the proposed adaptive algorithms in SCE
and DA schemes. We consider the scenario with a signal-to-noise
ratio (SNR) of $16$dB, $3$ users, and $1000$ training blocks.
Fig.\ref{fig:16dbblockberperfect} shows the BER performance of
different algorithms as a function of blocks transmitted. In this
experiment, the knowledge of the number of users $K$ and the noise
variance $\sigma_{e}^2$ are given for MMSE detection in the SCE
scheme. It will be shown later, the perfectly known $K$ and
$\sigma_{e}^2$ does not produce significant improvements in the BER
performance compared with using the estimated values. In both
schemes, with only $8$ iterations, the proposed CG algorithms
outperform the LMS algorithms and perform close to the RLS
algorithms. Since the filtering like step in the SCE scheme which
takes into the account that $L$ is smaller than $M$ provides some
performance gain, the adaptive algorithms in SCE scheme performs
better than in the DA scheme. However, the DA scheme has simpler
structure and lower computational complexity. The MMSE curves are
obtained with the knowledge of the channel, the spreading codes of
all the users and the noise variance. It can be found that, the MMSE
performances of the proposed schemes are exactly the same. This is
because these two schemes can be considered as two different
approaches to solve the same MMSE problem.

\begin{figure}[htb]
\begin{minipage}[b]{1.0\linewidth}
  \centering
  \centerline{\epsfig{figure=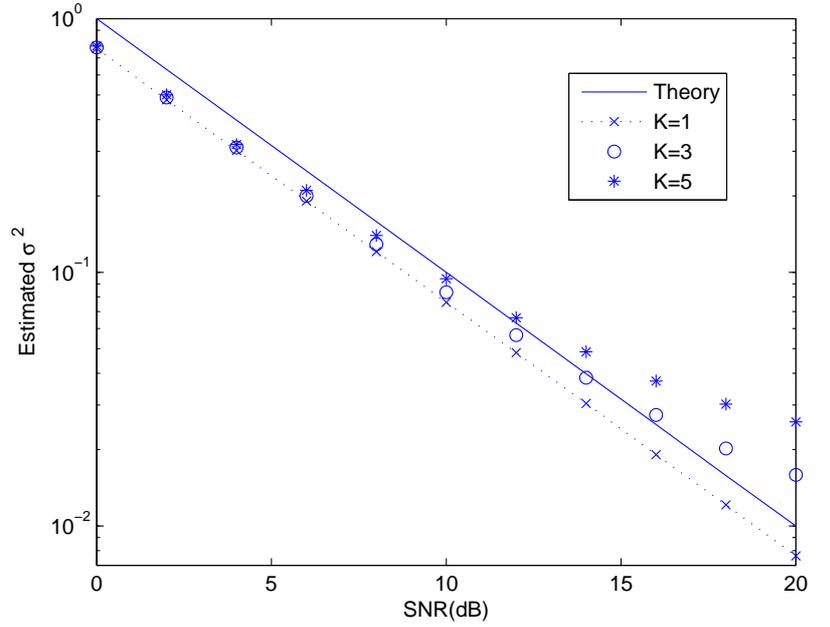,scale=0.8}}
\end{minipage}
\caption{Performance of the noise variance estimator.} \label{fig:sigma}
\end{figure}

Fig.\ref{fig:sigma} shows the performances of the ML estimators of
the noise variance in different SNRs. For each SNR scenario, the
estimated values of the noise variances for $1$, $3$ and $5$ users
are compared with the values in theory. For the multiuser case, the
ML estimators are not very accurate in the high SNR environments.
However, it will be demonstrated soon by simulations that this
inaccuracy will only lead to very limited BER performance reduction.

\begin{figure}[htb]
\begin{minipage}[b]{1.0\linewidth}
  \centering
  \centerline{\epsfig{figure=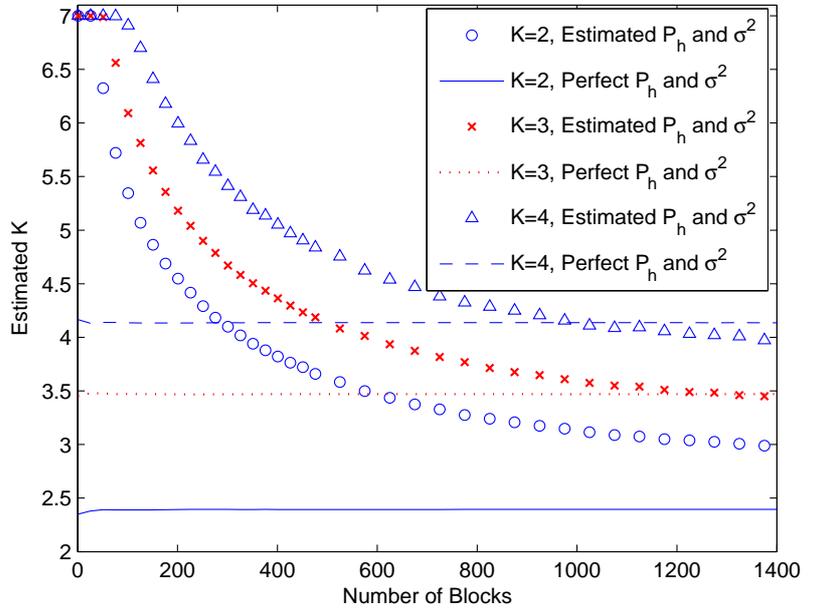,scale=0.8}}
\end{minipage}
\caption{Performance of the active users number estimator.}
\label{fig:kestiamte}
\end{figure}

Fig.\ref{fig:kestiamte} illustrates the performance of the
estimators of the number of active users in a $16dB$ environment
with SCE-CG algorithm and we consider the multiuser cases of 2 to 4
users. The number of users is determined by the received signal
power ${\rm P}_{\rm r}(i)$, the noise variance ${\sigma}^{2}_{e}(i)$
and the channel power ${\rm P}_{\rm h}(i)$ as shown in
\eqref{eq:forKfinalused}. Firstly, we show the performance of this
estimator with the knowledge of ${\sigma}^{2}_{e}(i)$ and ${\rm
P}_{\rm h}(i)$. Because of an approximation used in
\eqref{eq:forKfinalideal}, the values of the estimated users number
have gaps to the real values. For example, in 2 and 3 users cases,
these gaps are around 0.5 users. Secondly, we assess the performance
of the users number estimator with the estimated noise variance
$\hat{\sigma}^{2}_{e}(i)$ and the adaptive channel coefficients. It
should be noted that the channel estimation is started with a null
vector, which means very small $\hat {\rm P}_{\rm h}(i)$ at the
beginning stage and this leads to very large $\hat K$.  Hence, we
set $\hat K=7$ as an upper maximum for this estimator. The estimated
values of $\hat K$ approaches the curves which are obtained with the
knowledge of noise variance and the channel power. For 3 and 4 users
cases, the curves fit well but there is a mismatch when the users
number is 2. This mismatch is caused by the estimation errors of
$\hat{\sigma}^{2}_{e}(i)$ and $\hat {\rm P}_{\rm h}(i)$. However,
later simulations will indicate that the mismatches introduced by
the approximation and the estimation errors will not noticeably
affect the BER performance.

\begin{figure}[htb]
\begin{minipage}[b]{1.0\linewidth}
  \centering
  \centerline{\epsfig{figure=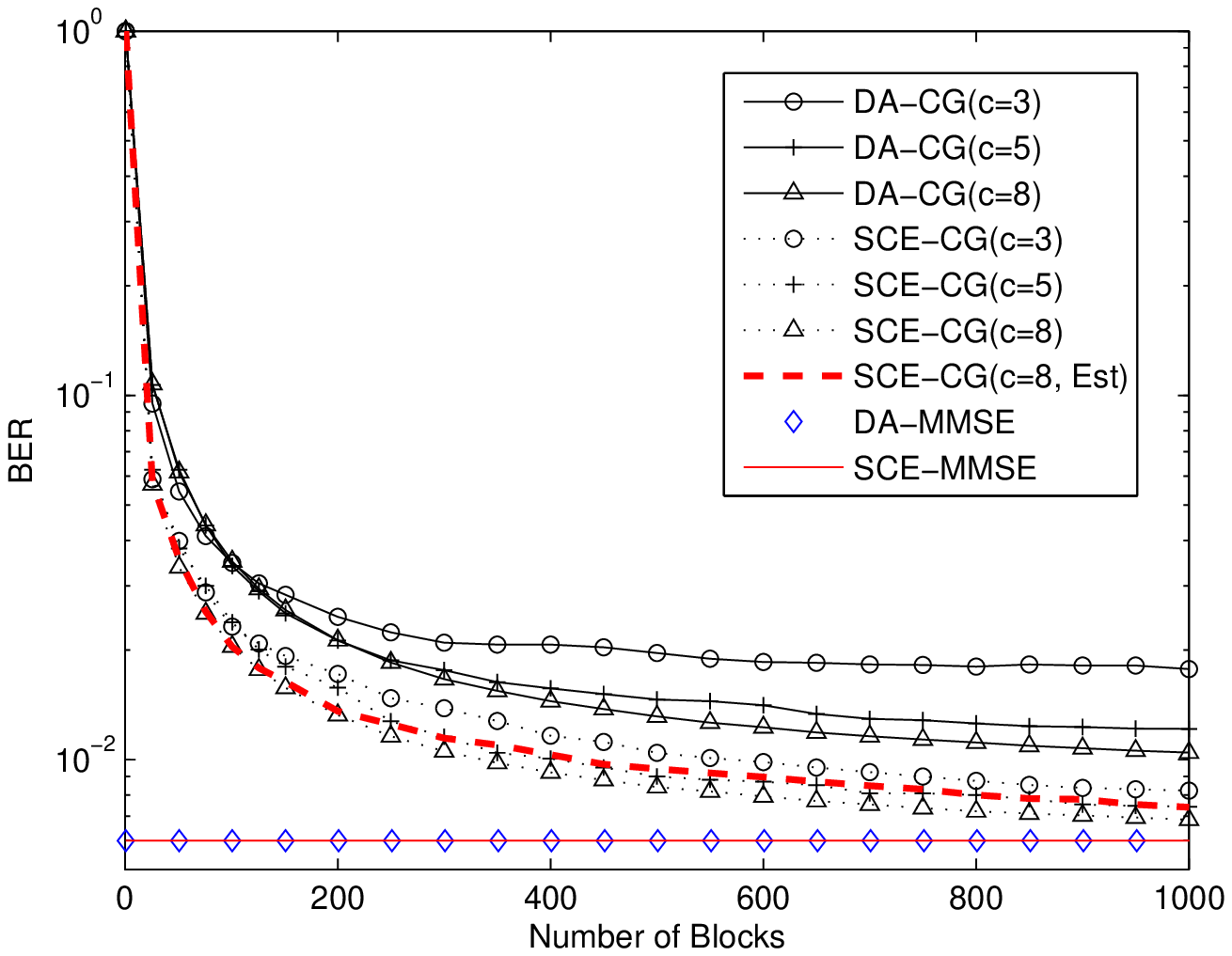,scale=0.8}}
\end{minipage}
\caption{BER performance of the proposed CG algorithms versus the number of
training blocks for a SNR=16dB. The number of users is 3.}
\label{fig:candestimate}
\end{figure}

Fig.\ref{fig:candestimate} shows the BER performance of the proposed
CG algorithms with different number of iterations for each
adaptation. For both schemes, the CG algorithms perform better as
the number of iterations increases. However, using more than $8$
iterations will only produce very limited improvement in the BER
performance for both schemes but increase significantly the
computational complexity. In our system, a good choice for the
number of iterations is $c=8$. In this figure, all the dotted curves
for the SCE scheme are obtained with the knowledge of
${\sigma}_{e}^{2}$ and $K$. We also include a dashed curve to show
the performance of the SCE-CG with 8 iterations that is using the
estimated values of $\hat{\sigma}_{e}^{2}$ and $\hat K$. It is shown
that using the estimated values will not affect the convergence rate
but only lead to a small reduction at the steady-state performance.

\begin{figure}[htb]
\begin{minipage}[b]{1.0\linewidth}
  \centering
  \centerline{\epsfig{figure=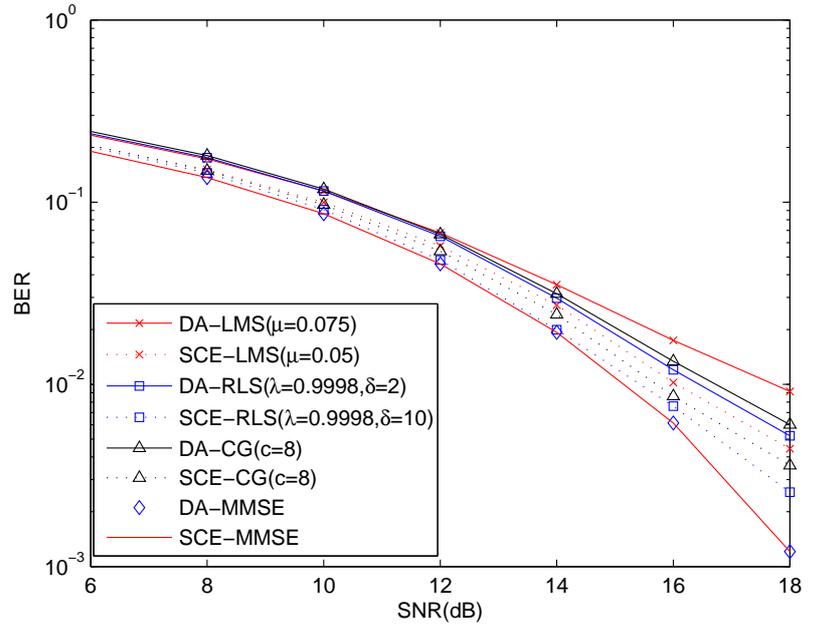,scale=0.8}}
\end{minipage}
\caption{BER performance of the proposed SC-FDE detection schemes versus the
SNR. The number of users is 3.} \label{fig:BERvsSNR}
\end{figure}

Fig.\ref{fig:BERvsSNR} illustrates the BER performance of different
algorithms in a scenario with $3$ users and different SNRs. In this
experiment, $500$ training blocks are transmitted at each SNR and
for the SCE scheme, the estimated $\hat{\sigma}_{e}^{2}$ and $\hat
K$ are used. For all the simulated SNRs, the proposed CG algorithms
outperform the LMS algorithms and are very close to the RLS
algorithms.

\begin{figure}[htb]
\begin{minipage}[b]{1.0\linewidth}
  \centering
  \centerline{\epsfig{figure=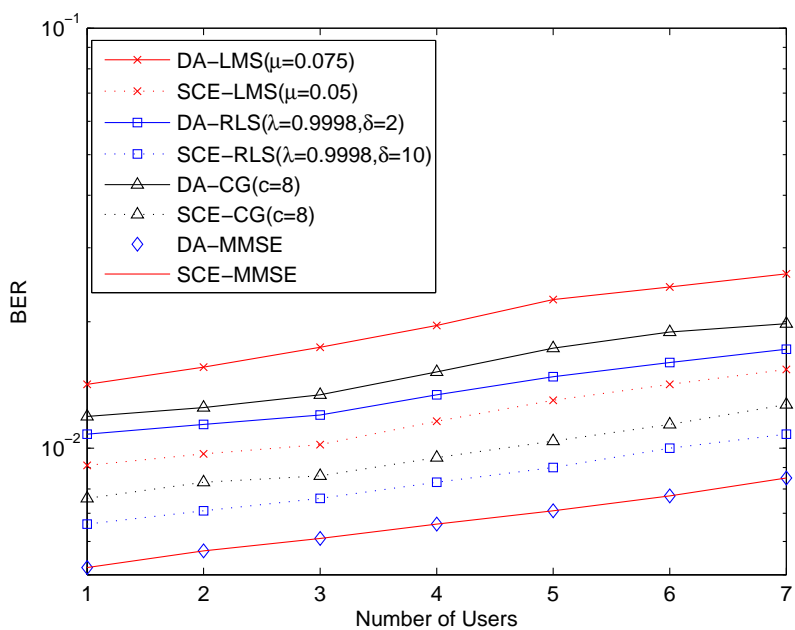,scale=0.8}}
\end{minipage}
\caption{BER performance of the proposed SC-FDE detection schemes versus number
of Users in the scenario with a 16dB SNR.} \label{fig:BERvsusers}
\end{figure}

Fig.\ref{fig:BERvsusers} shows the BER performance of different
algorithms in a $16dB$ scenario, with different numbers of active
users. The parameters for the adaptive algorithms are the same as
those used to obtain Fig.\ref{fig:BERvsSNR} and we use the estimated
values of $\hat{\sigma}_{e}^{2}$ and $\hat K$ for the SCE scheme.
For both schemes, the CG algorithms can support about $2$ additional
users in comparison with the LMS algorithms and the RLS algorithms
can support about $1$ additional user in comparison with the CG
algorithms.

\section{Conclusion}
\label{sec:conclusion} In this paper, two adaptive detection schemes
are proposed for the multiuser DS-UWB communications based on the
SC-FDE. These schemes can be considered as two approaches to solve
the MMSE detection problem in the block by block transmission SC
systems. The first scheme, named SCE, adaptively estimate the
channel coefficients in the frequency domain and then performs the
detection and despreading separately. In addition, the MMSE
detection in SCE scheme requires the knowledge of the noise variance
and the number of active users. To this purpose, we proposed simple
algorithms to estimate these values. The second scheme, named DA,
updates one filter in the frequency domain to suppress both the MAI
and the ISI. The DA scheme has simpler structure and lower
complexity but the SCE scheme performs better. For both schemes, we
developed LMS, RLS and CG adaptive algorithms. For the SCE scheme,
the CG algorithm has much lower complexity than the RLS algorithm
while performing better than the LMS algorithm. For the DA scheme, a
low complexity RLS algorithm is obtained which has the complexity
comparable to the CG algorithm in the multiuser scenarios. In the
single user system, the complexity of DA-RLS reduced to the same
level as the DA-LMS.


\newpage

\end{document}